\documentclass[12pt,preprint]{aastex}

\bibstyle{ApJ}
\usepackage{epsfig}
\usepackage{epsfig}
\usepackage{amsmath}
\usepackage{natbib}
\usepackage{subfigure}

\shorttitle{Physical Properties of Complex C Halo Clouds}
\shortauthors{Hsu, W.-H.}

\def\gtrapprox{\;\lower 0.5ex\hbox{$\buildrel >\over \sim\ $}}
\def\lessapprox{\;\lower 0.5ex\hbox{$\buildrel < \over \sim\ $}}

\def\Msun  {${\rm M}_\odot$}
\def\deg   {$^\circ$}
\def\arcmin {$^\prime$}

\def\NH    {$\rm N_{\scriptscriptstyle HI}$}
\def\HI    {H{$\rm{\scriptstyle I} $}}

\def\kms   {\ km s$^{-1}$}
\def\Mhi   {M$_{\rm HI}$}

\def\vgsr {$\rm V_{\rm GSR}$}
\def\vlsr {$\rm V_{\rm LSR}$}

\begin{document}

\title{Physical Properties of Complex C Halo Clouds}

\author{W.-H. Hsu\altaffilmark{1}, M. E. Putman\altaffilmark{2},  F. Heitsch\altaffilmark{3}, S. Stanimirovi{\'c}\altaffilmark{4}, J. E. G. Peek\altaffilmark{2}, S. E. Clark\altaffilmark{3}}
\altaffiltext{1}{Department of Astronomy, University of Michigan, Ann Arbor, MI 48109, USA; wenhsin@umich.edu}
\altaffiltext{2}{Department of Astronomy, Columbia University, New York, NY 10027, USA}
\altaffiltext{3}{Department of Physics \& Astronomy, University of North Carolina at Chapel Hill, Chapel Hill, NC 27599, USA}
\altaffiltext{4}{Department of Astronomy, University of Wisconsin - Madison, Madison, WI 53706-1582, USA}
\begin{abstract}

Observations from the Galactic Arecibo L-Band Feed Array \HI \ (GALFA-\HI)  Survey of the tail of Complex C are presented and the halo clouds associated with this complex cataloged.  The properties of the Complex C clouds are compared to clouds cataloged at the tail of the Magellanic Stream to provide insight into the origin and destruction mechanism of Complex C.  Magellanic Stream and Complex C clouds show similarities in their mass distributions (slope = -0.7 and -0.6 $\log(\rm{N}(\log(\rm{mass})))/\log(\rm{mass})$, respectively) and have a common linewidth of 20 - 30\kms \ (indicative of a warm component), which may indicate a common origin and/or physical process breaking down the clouds.  The clouds cataloged at the tail of Complex C extend over a mass range of $10^{1.1-4.8}$ \Msun, sizes of $10^{1.2-2.6}$ pc, and have a median volume density and pressure of 0.065 cm$^{-3}$ and (P/k) = 580 K cm$^{-3}$. We do not see a prominent two-phase structure in Complex C, possibly due to its low metallicity and inefficient cooling compared to other halo clouds.  From assuming the Complex C clouds are in pressure equilibrium with a hot halo medium, we find a median halo density of $5.8 \times 10^{-4}$ cm$^{-3}$, which given a constant distance of 10 kpc, is at a $z$-height of $\sim3$ kpc.  Using the same argument for the Stream results in a median halo density of $8.4 \times 10^{-5}~(60~\rm{kpc/d}) \ \rm{cm^{-3}}$. These densities are consistent with previous observational constraints and cosmological simulations. We also assess the derived cloud and halo properties with three dimensional grid simulations of halo \HI~clouds and find the temperature is generally consistent within a factor of 1.5 and the volume densities, pressures and halo densities are consistent within a factor of 3.

\end{abstract}

%Magellanic Clouds

\section{Introduction}
The origin and role of the neutral hydrogen (\HI) gas clouds located in the Galactic Halo is subject to much debate.  The suggested origin models include condensed gas originating from the ``Galactic fountain" \citep{SF76, Bregman80, FB08}, warm/hot halo gas cooling and fragmenting during the cooling process (e.g., \citealt{MB04,Kaufmann06, Sommer-Larsen06, Keres09}) and gas stripped from the dwarf galaxies (e.g., \citealt{Putman03}), with a combination of the models being most likely.  These halo clouds, a.k.a. high-velocity clouds (HVCs), potentially play a key role in galaxy evolution.   
Chemical evolution models of the Galaxy suggest that an infall of low-metallicity gas at the rate of $\sim 1$ \Msun \ per year is needed to explain the metallicity of the G and K stars in the solar neighborhood \citep{CMR01,FG03, Robitaille10} and our Galaxy is forming stars at a rate that seems to require continual re-fueling.  Galaxy formation simulations also suggest a galaxy gradually acquires its star formation fuel over time, and HVCs trace this ongoing process (e.g., \citealt{Peek08c, Keres09}).

Complex C and the Magellanic Stream are the largest and most massive high-velocity clouds in the Galactic sky (approximately \Mhi~= $5 \times 10^6$ and $2 \times 10^8$ \Msun \ respectively; \citealt{Thom08, Putman03}). While the Magellanic Stream has a known origin as being ripped from our dwarf companions, the Magellanic Clouds, Complex C's origin has remained a mystery.  
The argument that Complex C is a low metallicity infalling extragalactic cloud \citep{Wakker99} has been challenged by several subsequent authors, who find a metallicity range between 0.1 - 0.5 solar across the cloud  \citep{Gibson01, Collins03, Tripp03, Collins07}. This may indicate that Complex C is actually a mixture of infalling gas and ``Galactic fountain" gas, or possibly the stripped baryonic component of  a dark matter halo.  
%It has even been suggested that Complex C is part of an extended warp of our Galaxy (Kawata et al. 2003).
%In either case this cloud is representative of the origin and fate of high-velocity clouds as a whole.  

Complex C is elongated and extends from 140\deg \ - 30\deg \ in Galactic longitude and 65\deg \ -10\deg \ in Galactic latitude.   The tail of Complex C is therefore approaching the Galactic plane and may trace the accretion of halo gas onto the disk.  Recent distance constraints have opened up new possibilities for studies of the physical properties of Complex C (e.g., mass, physical size). \citet{Thom08} used halo stars to set upper and lower distance constraints on the complex and place it at 10 $\pm$ 2.5 kpc (see also \citet{Wakker07}).   They use this information to derive the total mass of the complex and estimate an accretion rate of 0.1 \Msun/yr from this HVC alone.

Both Complex C and the Magellanic Stream are thought to be embedded in a hot diffuse halo medium, as this halo medium is both detected indirectly observationally and expected from galaxy formation simulations.
The observational evidence for this halo medium includes: O VI absorption lines associated with the HVCs which are thought to originate from collisional ionization as the clouds interact with the ambient medium \citep{Sembach03};  head-tail structures of the HVCs, or a compressed head and diffuse tail (e.g., \citealt{Heitsch09,  Bruns00}); shearing structures on the \HI \ complexes \citep{Peek07}; confinement of the Magellanic Stream \citep{SDK02}; and the two-phase velocity structure of the clouds, implying that the HVCs lie in a medium of significant pressure (Wolfire et al. 1995b, hereafter W95; Kalberla \& Haud 2006\nocite{Wolfire95, KH06}).
Recent simulations indicate the hot halo medium fills the dark matter halo ($\sim150$ kpc) and hosts a large fraction of a galaxy's baryons (\citealt{MB04, Sommer-Larsen06, Kaufmann08}). This halo medium most likely originates from a combination of the initial baryon collapse and Galaxy feedback mechanisms, but its properties at a range of radii remain to be determined as it is extremely difficult to detect directly due to its hot, diffuse nature.  Detailed \HI \ observations of halo clouds at a range of distances can be used to probe this elusive, diffuse halo medium. 

In this paper we present new \HI \ observations of the tail of Complex C from the Galactic Arecibo L-band Feed Array \HI \ (GALFA-\HI) Survey (\S~2), catalog the clouds for this region and for the tail of the Magellanic Stream (\S~3), and derive the physical properties of the clouds (\S~4).  We test the derivation of some of these physical properties with simulations in \S~5.  Previous \HI \ observations of large sections of Complex C have been limited to observations with a 36\arcmin \ beam \citep{Wakker99, KH06}, while the GALFA-\HI \ observations provide 3.5\arcmin \ spatial resolution and up to 0.18\kms \ velocity resolution (smoothed to 1.4 \kms \ for this work).  The  GALFA-\HI \ observations of the Magellanic Stream were previously published \citep[][hereafter S08]{S08} and are cataloged and analyzed further here.   In \S~6, we compare the Complex C and Magellanic Stream cloud populations, investigate the physical underpinnings of the lack of two-phase structure in the clouds, and calculate the density of their surrounding diffuse hot halo through pressure-balance arguments.  The results provide insight into the physical properties of halo gas and the nature of HVCs as they disrupt within the halo. 

\section{Observations and Data Reduction}

\subsection{Observations}\label{obs}

The \HI \ data presented here were obtained with the GALFA-\HI \ spectrometer (galspect) on the Arecibo Radio Telescope \citep{SPH06}. GALFA-\HI \ data provide a channel spacing of 0.18\kms \ and cover a maximum velocity range of $-765$ to $+765$\kms \ (LSR).   The spatial resolution is approximately 3.5\arcmin.  The data presented here will be included in the GALFA-\HI \ Survey of the entire Arecibo sky and are publicly available at https:\/\/purcell.ssl.berkeley.edu.  

The Complex C data were taken as part of a proposal to map high and intermediate velocity clouds at the disk-halo interface (A2060). The observations were taken in ``basketweave mode". The telescope is pointed at the meridian but moves up and down in zenith angle. Each day the starting point was offset in RA so that the entire region of interest was covered. The advantage of this observing mode is that it generates a large number of crossing points which can be used in the crossing-point calibration (see Section~\ref{datareduction}). 

The Complex C map covers a region of 17h 20m to 18h 10m in RA and 2\deg \ to 14\deg \ in Decl., corresponding to approximately (l, b)= (25\deg \  - 40\deg, 12\deg \ - 24\deg,) in Galactic coordinates. We did two passes of this region to increase sensitivity and minimize the effects of interference.
The cube used to catalog clouds associated with the tail of the Magellanic Stream has been presented in S08.  
The Stream data cube is made up of both ``basketweave" and drift scans.  In this paper, we consider only the regions with the lowest noise level. The first region (referred to as region 1) extends from  22h 0m to 23h 30m in RA and 15.5\deg \ to 23\deg \ in Decl., corresponding to (l, b) = (77\deg \ - 97\deg, -41\deg \ - -28\deg) and the second region (referred to as region 2) extends from 22h 0m to 23h 50m in RA and 11.75\deg~to 16\deg~in Decl., corresponding to (l, b) = (72\deg \ - 100\deg, -48\deg \ - -32\deg).

\subsection{Data Reduction}\label{datareduction}

The data reduction includes the following steps, the details of which are outlined in \citet[][hereafter PH08]{Peek08}. 

1. Calibration using least-squares frequency switching: The least-squares frequency switching \citep{Heiles07} is used for GALFA-\HI \ observations because it is impossible to find an `off-target' region without Galactic emission near 0\kms \ that can be used in calibration. At the beginning of each observation, data from a single position are taken at several frequencies. This technique is able to separate the IF gain spectrum from the RF power spectrum, and thus the IF gain spectrum can be applied to all the spectra taken in that day.

2. Ripple removal:  Ripples due to known sources (reflection in the signal chain) are removed from the spectra. Since the geometry of the reflection is known, the Fourier components corresponding to these ripples can be easily removed. To remove the ripple caused by the reflection in the telescope superstructure and geodetic dome is more difficult. The average of all 7 beams is taken and subtracted from the average of one beam in a day. Then the resulting spectrum is searched for baseline ripples over the periods of $0.5 - 2 \mu s$.

3. Crossing-point calibration: At each crossing point, the sky is observed multiple times, and the dominating source of the difference in the spectra should be the variation in gain.  The crossing points can be used in determining the relative gains of each beam over each day.  The relative gain information is then applied to the calibration process and greatly reduces the effect of gain variation on the data. 

4. Gridding: The time-ordered spectra are gridded into the data cubes.
The GALFA-\HI \ Survey cubes are gridded into two formats, depending on the velocity region of interest and kinematic resolution required.   We chose to work with the cubes that have the entire GALFA-\HI \ velocity range and have the channels smoothed to 0.74\kms.   The cubes all have a 1\arcmin \ pixel size. 

The data used here have not been corrected for first sidelobes. Sidelobe calibration can be implemented in the future to improve the data quality \citep{Putman09}, and may increase the peak brightness temperatures of the clouds by at most 10\%, and decrease the cloud sizes by a similar amount. See section 6.6 in PH08 for details.

The rms noise level of the Complex C cube is about 0.06~K for channels smoothed to 1.4\kms. The 3-$\sigma$ sensitivity is ${4.5 \times 10^{17} \rm \ {cm^{-2}}}$ per 1.4\kms 
%\ or ${8.2 \times 10^{18} \rm \ {cm^{-2}}}$ per 25~\kms \ \citep[a linewidth typical for HVCs;][]{DBB02, KH06}. 
The 5-$\sigma$ mass sensitivity to a cloud at 10 kpc with 25 km/s linewidth is $\sim 18$ \Msun.   The rms noise level of the Stream cube is 0.05~K for region 1 and 0.03~K for region 2 for channels smoothed to 1.4~\kms. The 3-$\sigma$ sensitivity of the Stream cube is ${4.3 \times 10^{17}~\rm {cm^{-2}}}$ per 1.4~\kms \ for region 1 and ${2.3 \times 10^{17} \rm \ {cm^{-2}}}$ for region 2.  See S08 for further details.

\begin{figure}
	\begin{center}
	\includegraphics[scale=0.7,angle=-90]{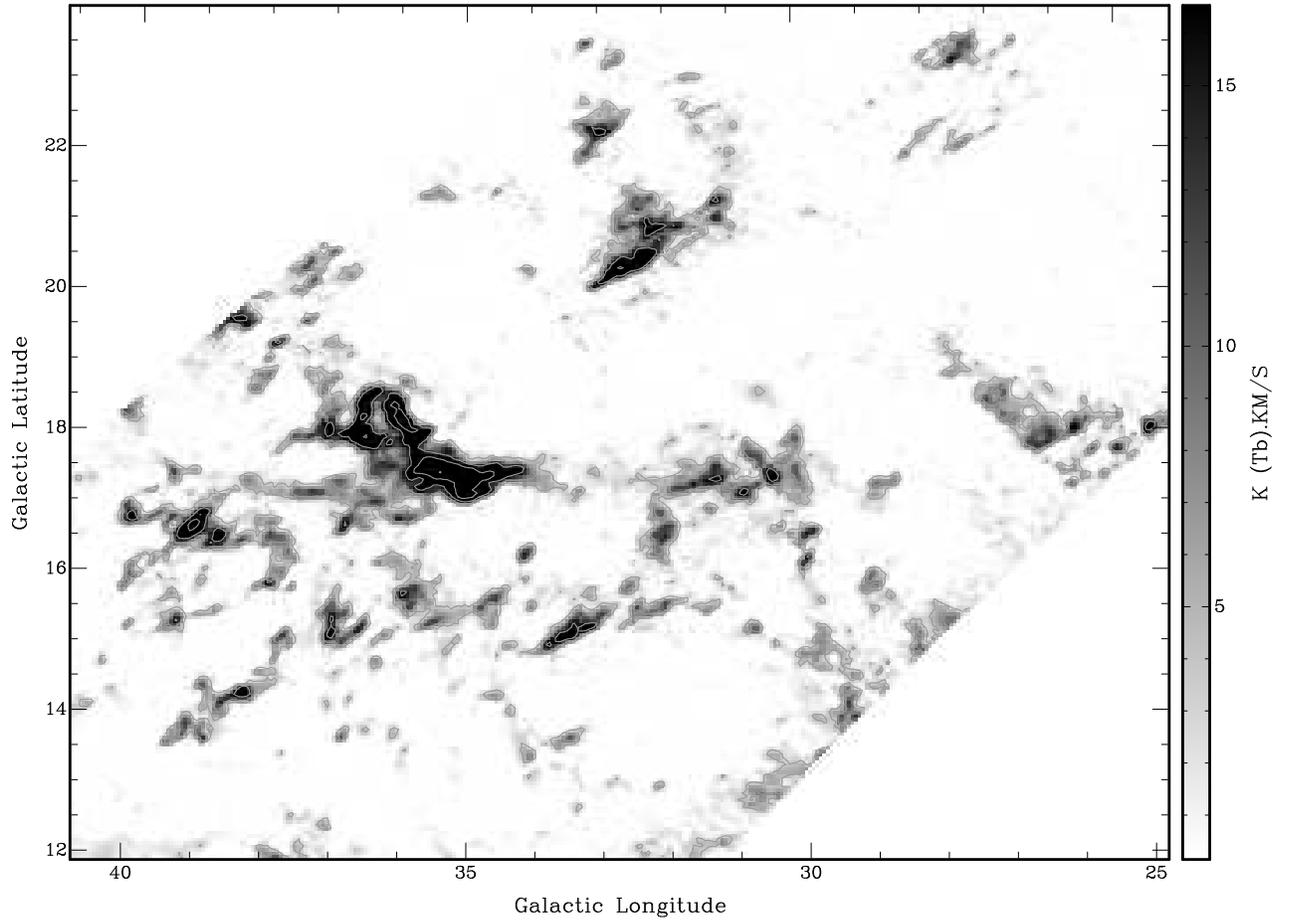}
	\caption{Integrated intensity map for the tail of Complex C in Galactic coordinates. The map covers \vlsr~= -190 to -65 \kms.  Contours are 4, 8, 16, 32 and 64 K~\kms, corresponding to column densities of $7.3\times10^{18}$, $1.5\times10^{19}$, $2.9\times10^{19}$, $5.8\times10^{19}$ and $1.2\times10^{20}$ cm$^{-2}$.}
	\label{moment0}
	\end{center}
\end{figure}

\begin{figure}
	\begin{center}
	\includegraphics[scale=0.7,angle=-90]{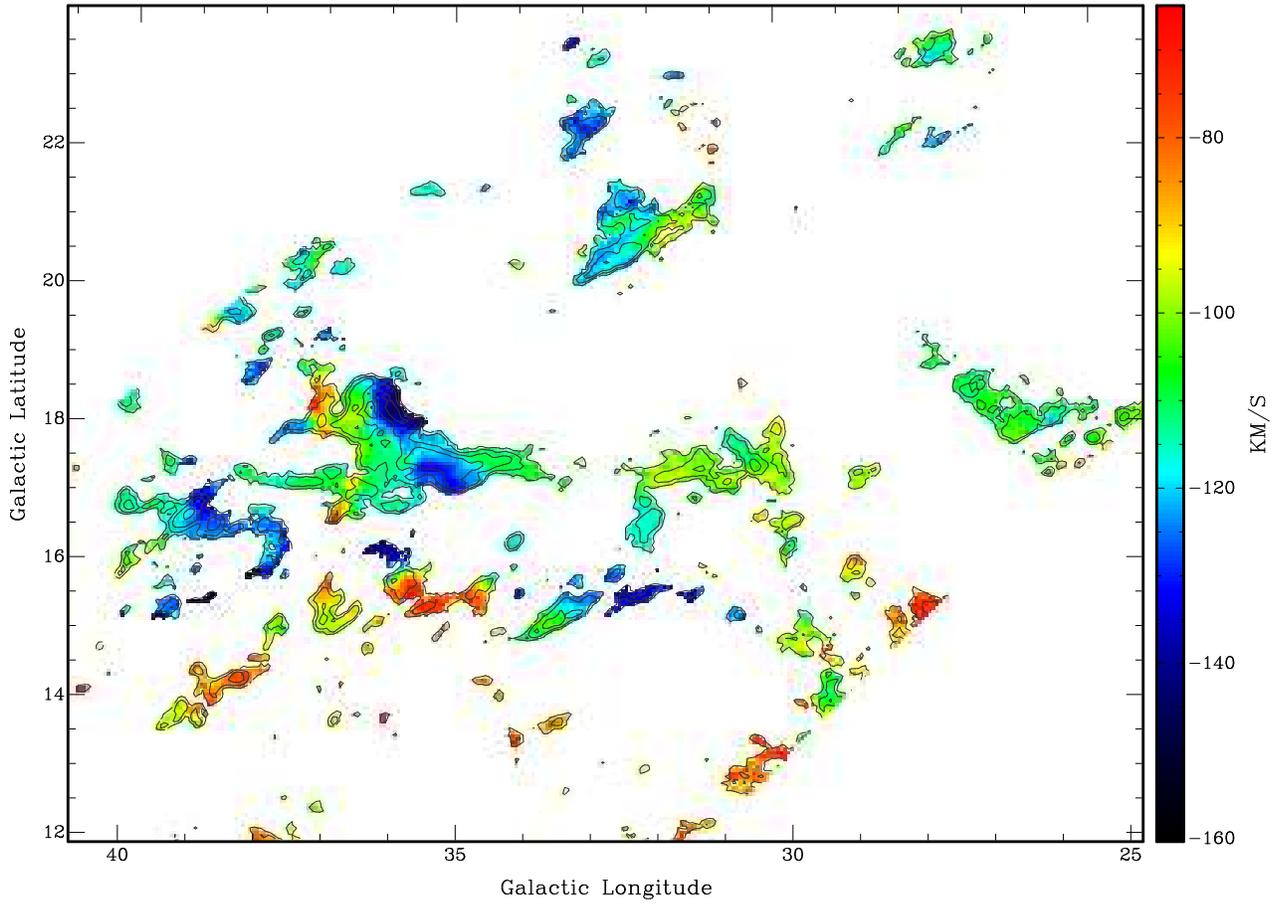}
	\caption{LSR velocity map of Complex C with the integrated intensity contours from Figure~\ref{moment0} overlaid. The map covers \vlsr~= -190 to -65\kms. A small fraction of clouds have \vlsr $ < $-160 \kms and are colored black in this map.}
	\label{moment1}
	\end{center}
\end{figure}

Figures \ref{moment0} and \ref{moment1} show the integrated intensity and average velocity maps of Complex C in Galactic coordinates, respectively.  These figures have been created after removing Galactic emission (described in Section~\ref{search}) and isolating the gas associated with Complex C (\vlsr~$-190$ to $-65$\kms as evident in the channel maps and cloud catalogs). 
%The maps include emission found over the velocity range of -190\kms \ to -65\kms, with the lower limit chosen based on where emission from the complex disappears in the data cube. 

\section{Cloud Catalogs}

To quantify the properties of individual clouds within the complexes we used an automated cloud finder called Duchamp (Whiting 2007).  This program searches for groups of connected voxels that are above a certain $T_b$ threshold and avoids bias introduced by cataloguing the clouds by eye.  Automating the cloud finding also allows us to quantify the depth of the search and how clouds are merged together.  This is particularly important for the results of this paper, giving us unbiased statistics of the cloud properties and allowing for a comparison of the clouds from two complexes.

%touchtodo
\subsection{Cloud Search $-$ Duchamp Source Finder}\label{search}

%Duchamp allows the user to specify selection parameters such as detection threshold and the minimum size of a detection.
We applied Duchamp to the GALFA-\HI \ cubes of the Magellanic Stream and Complex C as outlined below.  We also used Duchamp on a heavily spatially smoothed version of the Complex C cube to approximate observing the complex at a larger distance.  The steps used in the cloud cataloguing are summarized below, and the cataloguing parameters are summarized in Table \ref{tab1}.

1. The cubes are smoothed spatially to 3\arcmin \ by 3\arcmin \ per pixel and spectrally to 1.42\kms.  This increases the efficiency of the cloud finder by improving the S/N and having fewer pixels to search.

2. The robust statistics, the median and the median absolute deviation from the median (MADFM), are calculated from all the pixels and channels in the entire search region. The detection threshold for the cloud search is set at a fixed value n times the MADFM above the median ($n\sigma$) specified by the user. 

3. The data cube is searched for all pixels above the detection threshold. Once a pixel is found above the detection threshold, the size of the detection is increased by adding nearby pixels that are above a secondary threshold, or the 'grow parameter', specified by the user.

4. If two detections are adjacent to each other after step 3, the detections are merged into one.

5. All the detections from step 4 are automatically screened by Duchamp for false detections.  To minimize this number, we specify that the detections must span a minimum of four channels (5.6 \kms) and two pixels ($\approx 6$\arcmin) to be included in the output list.

6. All the detections in the output list are then examined by eye. Detections that are possibly due to scanning artifacts or RFI are excluded from the final list.

The output of Duchamp provides the following information: RA and Dec of the cloud, size of the cloud in RA and Dec, and the total spectrum of all the detected pixels. A Gaussian fitting program was subsequently used to fit up to two Gaussian profiles to the spectrum of each cloud.  We only adopt the result of the two-Gaussian fit if it improves the rms error by more than 5\%.  The linewidth, central velocity, and total \HI \ column density of the cloud were determined from this Gaussian fitting. 
The parameters used to search the Complex C and Magellanic Stream cubes and the cataloguing results are summarized in the next two sections.

\begin{figure}
	\begin{center}

\subfigure[Spectrum of cloud 35 before removing Galactic emission]{\includegraphics[bb=60 645 430 733,clip]{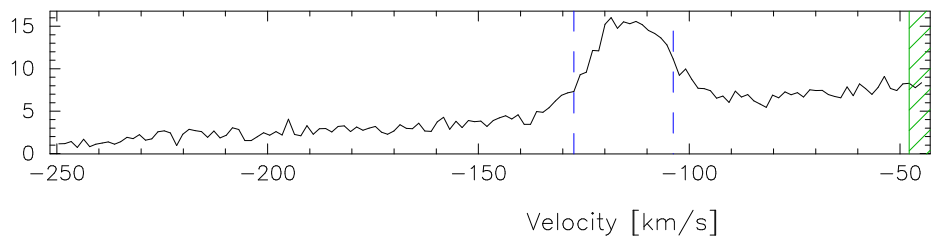}}
\subfigure[Spectrum of cloud 35 after removing Galactic emission]{\includegraphics[bb=60 645 430 733,clip]{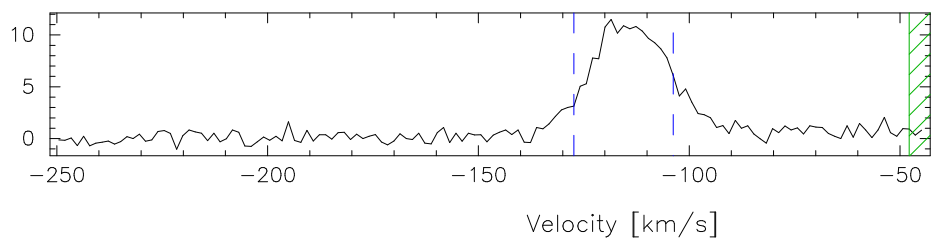}}
\subfigure[Spectrum of cloud 77 before removing Galactic emission]{\includegraphics[bb=60 645 430 733,clip]{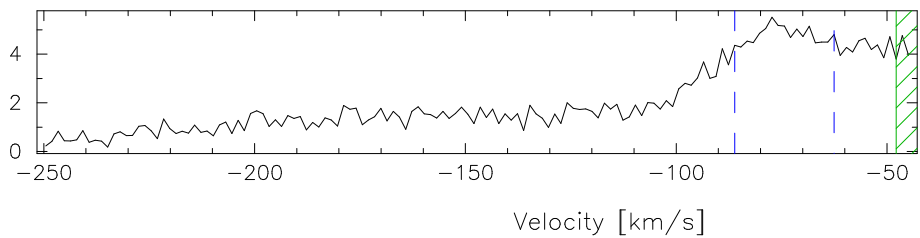}}
\subfigure[Spectrum of cloud 77 after removing Galactic emission]{\includegraphics[bb=60 645 430 733,clip]{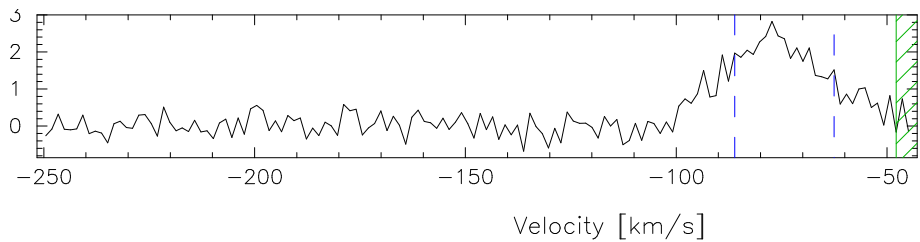}}
\caption{Two examples of the integrated spectra of cataloged Complex C clouds before and after removing Galactic emission (labeled 35 and 77 in Figure~\ref{Cmom}). In the case of cloud 35, the cloud would be easily identified by Duchamp without Galactic removal (but with an elevated flux). In the case of cloud 77, removing Galactic emission makes it much easier to identify the cloud.  The units on the y-axis of the spectra represent the sum of the brightness temperatures of all pixels associated with cloud and the hatching to right represents the velocity region not considered in the cataloging due to stronger Galactic emission.}
%\caption{Example of the spectrum of a cataloged Complex C cloud (labeled 35 in Figure~\ref{Cmom}). The top figure shows the integrated spectrum before removing Galactic emission; the bottom is the integrated spectrum of the same cloud after removing Galactic emission.}
\label{Cexamp}
	\end{center}
\end{figure}

There was an additional step to the above list in the case of Complex C.  Since Complex C extends into less negative velocities where Galactic emission becomes important, the noise level in the Complex C data is not constant throughout the spectrum. It is difficult to use Duchamp directly on the cube in this case because Duchamp only uses the global statistics of the cube.  To solve this problem, we made an effort to remove Galactic emission from the cube. 
Since diffuse Galactic emission does not change dramatically over the spatial region of the Complex C cube, we determine the contamination of Galactic emission by using a few small patches where there are no discrete \HI \ clouds.   The size of each patch is 30\arcmin~$\times$ 30\arcmin, and the patches are roughly 3 - 4 degrees apart.  The distance weighted average spectrum of these patches is then subtracted from the -250 to -40\kms~range of each spectrum in the cube. 
This step is done after the cube is smoothed into 3\arcmin~$\times$ 3\arcmin \ pixels and before cloud searching (i.e., between steps 1 and 2). This made it possible to catalog discrete clouds at low velocity and also gives a more accurate mass for these clouds. Figure~\ref{Cexamp} shows examples of cloud spectra before and after removing Galactic emission.

\subsection{Complex C Cloud Catalog}

\begin{figure}
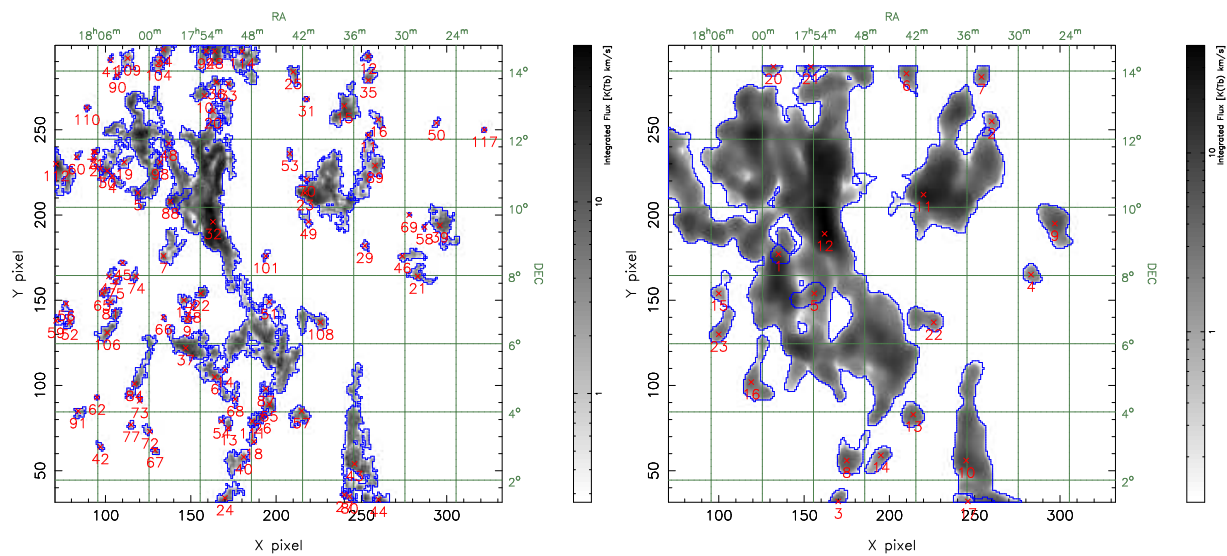

\includegraphics[bb=45 0 550 490,clip,scale=0.45]{f4a.ps}
\includegraphics[bb=45 0 550 490,clip,scale=0.45]{f4b.ps}
\caption{The intensity maps of the detections in the Complex C datacube as provided by Duchamp. Note that some numbers are missing as they are excluded from the catalog (see Sec~\ref{search}).  The left map has the original spatial resolution of 3.4\arcmin. Clouds 1 to 90 are included in Table~\ref{Ctab}, and clouds 91 to 117 have \vlsr~$> -65$\kms~and are included in Table~\ref{Ctab2}. The right map is spatially smoothed to 18\arcmin\ before the clouds are cataloged. Clouds 1 to 16 are cataloged in Table~\ref{CStab}, and clouds 17 to 23 have \vlsr $~> -65$\kms~and are not cataloged.}
\label{Cmom}
\end{figure}

We ran Duchamp twice on the Complex C cube; once at the full resolution (3.5\arcmin) and once smoothed to a resolution equivalent of moving Complex C to a distance more appropriate for the Magellanic Stream (60 kpc, 18\arcmin).  We used a detection threshold of 0.29 K (5$\sigma$) on the original cube and 0.12 K (5$\sigma$) on the smoothed cube, and grew the detected clouds to 3$\sigma$ (0.18 K on the original cube, and 0.08 K on the smoothed cube). The search parameters are tabulated in Table~\ref{tab1}. The velocity range searched for clouds was -200 to -50\kms.  Figure~\ref{Cmom} shows the integrated intensity maps of the Duchamp detections for the original Complex C cube and the smoothed cube.  We inspected each of the Duchamp detections by eye after running the program (step 6).  About 15\% of the Duchamp detections (20 out of 134) were removed due to their identification as remnant scanning artifacts (detections which generally only extend over a few pixels and/or channels or obviously follow the basket weave pattern) and an additional 21 clouds were removed because a large percentage of their emission was located at the edge of our search area (i.e., velocity center lies near or above -50 \kms).   An example spectrum of a real Duchamp detection in Complex C is shown in Fig~\ref{Cexamp}.  Only the real clouds with \vlsr~$< -65$\kms~are cataloged in Table~\ref{Ctab}.  Clouds between \vlsr~= -65 to -50 \kms~are cataloged in Table~\ref{Ctab2}.   
%This break in velocity is explained in \S~4.1.1.

\begin{figure}
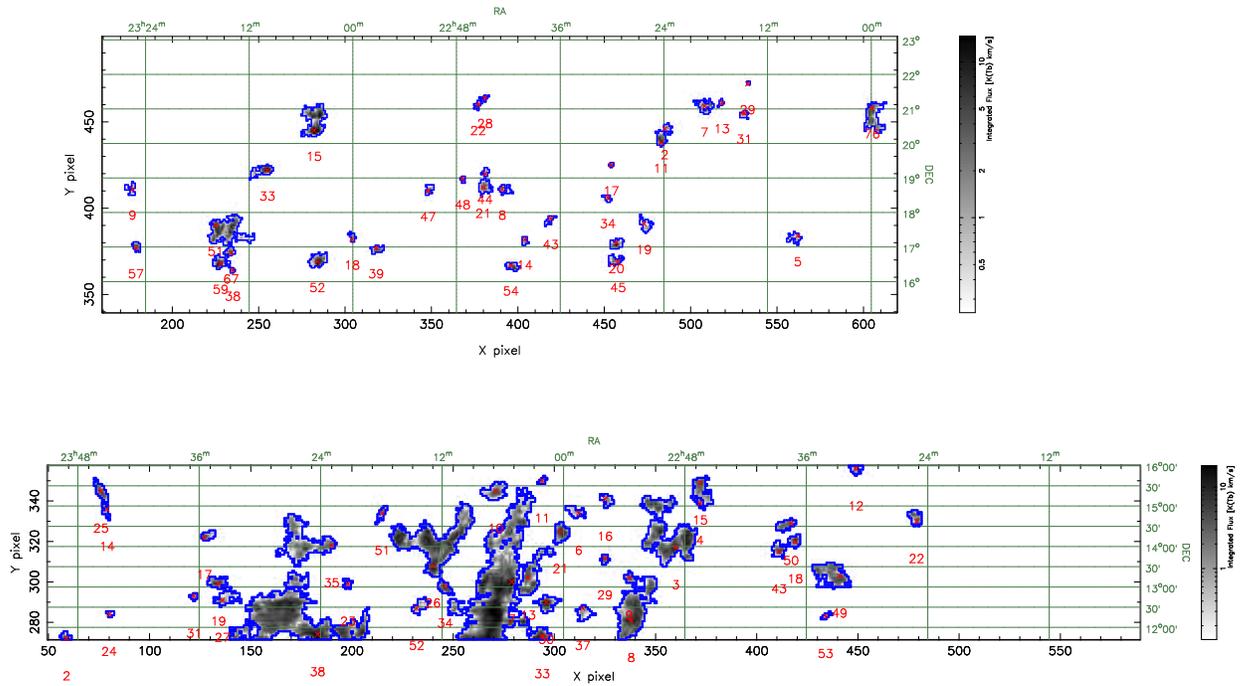

	\begin{center}
	\includegraphics[bb=45 25 550 250, clip, scale=0.8]{f5a.ps}
	\includegraphics[bb=45 0 550 150, clip,scale=1.1]{f5b.ps}
	\label{MSmom}
	\caption{Integrated intensity map of the detections in two regions at the tail of the Magellanic Stream as provided by Duchamp (top corresponds to ``region 1" and bottom corresponds to ``region 2"). See Table~\ref{Stab} for the data on each numbered cloud.}
	\end{center}
\end{figure}

The cloud catalogs contain the following entries:

Column 1: Cloud number. Here we use the cloud numbers assigned by Duchamp  (shown in Fig~\ref{Cmom}, Fig~\ref{MSmom} for the Magellanic Stream). The missing numbers are clouds removed after examining each Duchamp detection by eye.

Column 2 and 3: RA and Dec (J2000) as provided by Duchamp. These are the intensity-weighted average centroid positions.

Column 4 and 5: Galactic longitude and latitude obtained from the RA and Dec in Column 2 and 3.

Column 6 and 7: Angular size in RA and Dec. The angular size is given by Duchamp and is the total size of the cloud down to 3$\sigma$ level. At the distance of 10 kpc, the conversion between angular size and physical size is given by 1\arcmin~= 2.9 pc.

Column 8: Peak brightness temperature in K.

Column 9: Local Standard of Rest (LSR) velocity.  Centroid determined by Gaussian fitting of the total cloud spectrum.

Column 10: Velocity in the Galactic Standard of Rest frame (GSR), defined as \\ \vgsr~$ = 220 \cdot \cos(b)~\sin(l)~+~$\vlsr.

Column 11: Width of the Gaussian used to fit the spectrum in \kms, defined as the FWHM of the Gaussian.  

%Column 11: Integrated intensity of the total cloud, in units of K\kms.

Column 12: Total \HI \ column density of the cloud, in units of $10^{18}\; \rm cm^{-2}$.

Column 13: \HI \ mass (in \Msun), assuming all observed Complex C clouds are 10 kpc away (Thom et al. 2008) as described in Sec~\ref{results}.

Table~\ref{Ctab} (\vlsr~$< -65$\kms) contains 79 clouds.
If two Gaussians are required to fit the total spectrum of a cloud, the two Gaussians will be listed separately in the catalog. The two rows will share the same position and size information (columns 1 to 7), but have different velocities, \HI \ column densities and masses (columns 8 to 12).  The clouds between \vlsr~= -65 to -50\kms~are included in a separate catalog containing 14 clouds (Table~\ref{Ctab2}).  This separate catalog was created because we found a break in the velocity distribution of the clouds at approximately -65~\kms, indicating this is a natural cut-off point for clouds that are clearly associated with Complex C.   The clouds of Table~\ref{Ctab2} are also not included in the statistics discussed in \S~\ref{results}.

In the smoothed Complex C cube, the number of detected clouds is much smaller (16) because small discrete clouds are often merged into larger clouds.  Table~\ref{CStab} shows the catalog of clouds created from the smoothed cube, where we exclude the clouds that have a \vlsr~between -65 and -50\kms.  

\subsection{Magellanic Stream Cloud Catalog}

The Magellanic Stream data have different noise levels in the two spatial regions considered here due to differing integration times (see S08). 
The statistics are determined separately for these two regions: for region 1 we used a detection threshold of 0.27 K (5$\sigma$) and a grow parameter of 0.17 K (3$\sigma$); for region 2 we used a detection threshold of 0.15 K (5$\sigma$) and a grow parameter of 0.09 K (3$\sigma$).  We searched the cubes from -420 to -280~\kms, where there is evident Stream emission.  The search parameters are tabulated in Table~\ref{tab1}.

 The Stream cubes are plagued with more RFI than the Complex C cube.  We excluded 62\% of the clouds found by Duchamp in region 1 and 31\% of the clouds in region 2 due to RFI artifacts or overlap of the two regions.  Figure \ref{MSmom} shows the integrated intensity map of the real cloud detections and Table~\ref{Stab} contains the catalog of clouds found in the Magellanic Stream data. The entries are the same as those in Table~\ref{Ctab} with the exception of the \HI \ mass being calculated at a distance of 60 kpc.  Also, at the distance of 60 kpc, the conversion between angular size and physical size is given by 1\arcmin~= 17 pc.  These relationships can be easily scaled to other distances using \Mhi(d (kpc)) =\Mhi$\rm{(60~kpc)(d/60~kpc)}^2$ and $\rm{size(60~kpc)}(d/60~kpc)$ or 17~pc $\times$ size(\arcmin)$\rm{(d/60~kpc)}$.

\section{Results}\label{results}

In this section, we present the distributions of cloud properties and, given we have distance constraints for Complex C and the Magellanic Stream, derive the physical properties of the clouds.  A summary of the cloud properties is found in Table~\ref{average}.  The physical properties are derived using the following methods.

\begin{itemize}
\item The mass of a cloud is derived at the distance of 10 kpc in the case of Complex C and 60 kpc for the Magellanic Stream from the total column density of the cloud using, \Mhi(\Msun)$=5.5 \times 10^{-21}$ \NH (tot)$\ \times \ \rm{d(kpc)}^2$.
%at a given distance can be calculated from its total flux by using \Mhi (\Msun)${=1.26 \times 10^{-2} \rm {\ I(K}}$\kms ${)\times \rm {d(kpc)}^{2}}$, where I is the integrated intensity and d is the distance to the object. This is derived by taking into account of the total column density of the cloud and the beam size at the distance of the cloud.

\item The angular size of a cloud is given by ${\sqrt{\rm{\Delta RA \; \Delta Dec}}}$, and the physical size $= 0.291 \times \rm{angular~size}~\times$ d(kpc). 

\item The volume density (n${_{c}}$) of a cloud is calculated by assuming that the clouds are spherical, with the radius R = ${\sqrt{ \rm{\Delta RA \; \Delta Dec}}/2}$.  Clouds with aspect ratios greater than 1.6 are excluded from this calculation.

\item The cloud pressure is given by $\rm{P_{c}=k n_{c} T_{c}}$, where we assume the warm neutral component is at 9000K. 
\end{itemize}

\subsection{Complex C}

\subsubsection{The Original Complex C Cube}

79 clouds were cataloged in the searched region of the Complex C cube at \vlsr \ $< -65$ \kms, and 14 were cataloged at $-65 < \ $\vlsr$ \ <-50$\kms. Traditionally high-velocity clouds such as Complex C are defined to have \vlsr $\lessapprox -90$\kms, but from the distribution of clouds in the data cube, there is no distinct cutoff at this velocity.  There is however, a minimum number of clouds around -65\kms, and the channel maps also show a transition from Complex C to a separate, intermediate velocity population around this velocity. Thus in this analysis we only include the clouds with \vlsr$ \ <-65$\kms. We inspected the clouds at $-65 < \ $\vlsr$ \ <-50$\kms, and found that the overall statistics we present here would not change significantly if we included these clouds.  They are included in Table~\ref{Ctab2} for those interested in that set of clouds.

Figure~\ref{C_results1} shows the peak $T_b$, central velocity, linewidth, and angular size for the 79 clouds cataloged.  The distance dependent properties of the clouds are shown in Figures 8 \& 13 with the exception of the physical size of the clouds at 10 kpc, which for ease of presentation we show as the top axis on the angular size distribution plot.
Most of the clouds have a peak $T_b$ under 1 K with a few exceptions.  The cutoff at 0.29 K is determined by the noise level of the cube and the cataloging parameters, and there is no evidence for a turnover before this value, which suggests that the cloud number continues to increase below our detection limit. The linewidth histogram shows that the cloud linewidth distribution peaks between 20 - 30\kms, 67\% of the clouds are found within this peak. %87\% of the clouds have linewidths between 15 and 35\kms. 
The median linewidth is 24.9 \kms. 8 of the clouds required a double Gaussian to fit to their line profiles. The solid lines in the linewidth and central velocity plots show only the clouds that were fit with a single Gaussian, while the dashed lines include the clouds fit with either one or two Gaussians, treating the two components as two separate clouds. 

The velocity distribution of the clouds extends from -178 to -64\kms~(LSR; 41 - 84 \kms~in the GSR frame), with the upper cutoff described above.   The distribution has a peak at -120\kms, but it is not very pronounced.   There is a steady decline in the number of clouds as one approaches less negative velocities.  
%, and then the number of clouds increases again at around -55\kms.  This increase in the number of clouds makes sense as emission from the Galactic Plane is being approached. 
The sizes of the clouds are shown in both arc minutes and the corresponding physical size at 10 kpc. 66\% of the clouds have sizes of 10\arcmin \ to 30\arcmin, corresponding to physical sizes of 30 to 100 pc. The median cloud size is 22\arcmin, corresponding to a physical size of 64.4 pc. 

Figure~\ref{C_results2} shows some of the physical properties of the clouds that depend on distance.  The first two plots show the mass and size of the clouds vs. linewidth. Clouds fit with double Gaussians are excluded from these plots. The relation between log(mass) and log(linewidth) is not very pronounced but the trend is that the larger the mass, the larger the linewidth. The slope of the linear fit is 0.08 $\pm$ 0.02 $\log(\rm{linewidth})/\log(mass)$. The relation between log(size) and log(linewidth) also shows a general trend that a larger size corresponds to a larger linewidth.  The slope of the linear fit is 0.14 $\pm$ 0.03 $\log(\rm{linewidth})/\log(\rm{size})$.  Given the selection effect of small and/or low-mass clouds with large linewidths being difficult to detect, measurement of this slope should not be considered a significant result.

The bottom left panel shows the distribution of \HI \ masses at 10 kpc for the cataloged clouds.  The
% integrated intensity of the complex C clouds ranges from ${10^{1.1}}$ to ${10^{4.8}}$ K\kms, which corresponds to a 
mass range extends from ${10^{1.1}}$ to ${10^{4.8}}$ \Msun. (Because there is only one cloud in the last bin, it is not shown in this plot.)  The mass distribution (given they are all placed at the same distance) follows a power law of slope of -0.60 $\pm$ 0.05 $\log(\rm{N}(\log(\rm{mass})))/\log(\rm{mass})$.  The cutoff on the low end is determined by the noise level and the cutoff at the high end may be partially due to the limited spatial size of the cube.  The drop in the number of clouds below ${10^{1.6}}$ \Msun~is most likely due to catalog completeness and sensitivity limitations.  In the case where the clouds were fit with two Gaussians, the sum of the two Gaussian components is used for the mass. The total \HI \ mass of the clouds is ${10^{5.0}}$ \Msun. The mass of the cataloged clouds is about 2\% of the total mass of Complex C (approximately \Mhi~= $5 \times 10^6$~\Msun, \citealt{Thom08}). 

The volume density distribution of the clouds is shown in Figure~\ref{halo_density_fig} (upper left). Since we do not have information on the size of the cloud in the third dimension, the clouds are assumed to be spherically symmetric, and the geometric mean of the size in RA and Dec is used for the diameter. Clouds that have an aspect ratio of greater than 1.6 are excluded from the density plots because they clearly violate the spherical symmetry assumption. The volume density ranges from 0.002 to 0.35 cm${^{-3}}$, but the distribution has a prominent peak around 0.06 cm${^{-3}}$, with a median volume density of 0.0645 cm$^{-3}$. About 60\% of the clouds have a density between 0.01 to 0.08 cm${^{-3}}$.  The distribution of derived cloud pressures is shown in the lower left plot of Figure~\ref{halo_density_fig} (bottom axes), with most (59\%) of the values being $P/k = 10^{2.6-3.1}$ K cm$^{-3}$ and the median value being $10^{2.8}$ K cm$^{-3}$. 

\subsubsection {Smoothed Complex C clouds}
The search on the smoothed Complex C cube yields significantly fewer clouds as some small clouds are merged into larger clouds and some fall below the 5$\sigma$ threshold after smoothing. Only 16 clouds are cataloged, and 2 of them required double Gaussian fitting to their total spectrum. Only clouds with \vlsr $\ <-65$\kms \ are cataloged and included in the statistics.

Figure \ref{CCS_results1} shows the peak $T_b$, central velocity, linewidth, angular size and corresponding physical size at 10 kpc for all of the cataloged clouds from the smoothed cube.  The shape of the peak $T_b$ distribution is similar to that of the original cube, but the values are significantly smaller. This makes sense because when the cube is smoothed, the $T_b$ of the peak pixel is distributed to other pixels. The linewidth histogram has a peak at a similar velocity range as the full resolution catalog (24 - 28\kms), with most clouds (67\%) within 19 - 30\kms~and a median value of 26.2 \kms. The velocity distribution of the clouds in the smoothed Complex C cube extends from -130 to -80\kms. The few clouds at the highest velocities ($\sim-170$\kms) in the original cube are merged into clouds of much higher flux at lower velocities and thus become a faint tail of high velocity emission for a much larger cloud.   Most of the clouds are between -120 and -100\kms.   The size distribution of the clouds is significantly larger, with the peak dictated by the spatial smoothing of the cube as expected. The median cloud size in the smoothed Complex C cube is 51.7\arcmin, corresponding to 150 pc at the distance of 10 kpc.

Figure~\ref{CCS_results2} shows the mass and size of the clouds vs. linewidth and the \HI~mass at 10 kpc.  The relation between log(mass) and log(linewidth) is statistically indistinguishable from that of the original cube, though it is obviously greatly affected by low number statistics. 
The log(size) vs. log(linewidth) plot is also statistically indistinguishable from that of the original cube. 
Since the smaller clouds are merged into larger clouds, the total \HI \ masses are in general larger. The 
mass range of the clouds is ${10^{1.9}}$ to ${10^{4.9}}$ \Msun.   Again the most massive cloud is not shown on this plot and for clouds fit with two Gaussians, the sum of the two Gaussians was used for the integrated intensity and mass.  With the small number of clouds, it is not possible to determine if this distribution follows a power law.  The total \HI \ mass of the clouds is the same, ${10^{5.0}}$ \Msun.  The volume density of clouds in the smoothed cube is shifted to smaller values, though a very limited number of clouds satisfy our aspect ratio criteria (13).  This is expected as many clouds are merged together in the smoothing.  

\subsection{Magellanic Stream}
68 clouds were cataloged in the searched cubes containing the Magellanic Stream. 34 clouds were found in region 1, and 38 were found in region 2 (see Section~\ref{obs} or Figure~\ref{MSmom} for the regions), with only 4 of the clouds requiring double Gaussian fitting.  S08 also catalog clouds
in these regions, but does so by eye.  They find only 9 clouds in region 1 and 55 clouds in region 2.  The discrepancies are due to two reasons: Duchamp is able to find smaller and fainter clouds that might be omitted in by-eye searches, and it counts two clouds as one if they are connected by diffuse emission that is above the grow threshold.

Figure~\ref{MS_results1} shows the peak $T_b$, central velocity, linewidth and size for all of the clouds cataloged. More than 90\% of the clouds have a peak $T_b$ less than 0.6 K.  The detection limit is 0.27 K for region 1, and 0.15 K for region 2.  There is no evidence of a turnover in the distribution.
The linewidth histogram spans from 10 to 65\kms, but 75\% of the clouds are found between 20 - 35\kms, with a median linewidth of 27.7 \kms. 
The velocity distribution of the clouds in the Stream cube extends from -420 to -300\kms~(LSR), with a peak at -350\kms.  In the GSR frame this range corresponds to -247 to -115 \kms.
The sizes are shown in both arc minutes and the corresponding physical size at 60 kpc. The sizes of the clouds range from 6\arcmin \ to 170\arcmin.  70\% of the clouds have sizes of 10\arcmin\ to 30\arcmin, which corresponds to 170 to 510 pc at the assumed distance of 60 kpc. The median cloud size is 22.1\arcmin, corresponding to 376 pc (d/60~kpc).

The two upper plots of Figure~\ref{MS_results2} show the mass and size of the clouds vs. linewidth.  The relation between log(mass) and log(linewidth) is not very pronounced but the trend is that the larger the mass, the larger the linewidth. The slope of the linear fit is 0.12 $\pm$ 0.02 $\log(\rm{linewidth})/\log(mass)$. The relation between log(size) and log(linewidth) also shows a general trend that the larger the size is, the larger the linewidth is. The slope of the linear fit is 0.23 $\pm$ 0.04 $\log(\rm{linewidth})/\log(\rm{size})$.  As noted previously, selection effects are not taken into account for these slopes.  The lower left corner of Figure~\ref{MS_results2}  shows the distribution of \HI \ masses at 60 kpc for all the cataloged clouds.  
The clouds extend over a mass range of ${10^{2.4}}$ to ${10^{5.5}}$ \Msun, and follow a power law of slope -0.70 $\pm$ 0.03 $\log(\rm{N}(\log(\rm{mass})))/\log(\rm{mass})$. The total \HI~mass of the clouds is $10^{6.1}$ \Msun $\rm{(d/60~kpc)}^2$.

The upper right corner of Figure~\ref{halo_density_fig} shows the volume density distribution of the clouds if they are located at 60 kpc. The density ranges from 0.001 to 0.03 cm${^{-3}}$, and 63\% of the clouds have a density between 0.003 to 0.012 cm${^{-3}}$. The median volume density is 0.009 cm$^{-3}$ $\rm{(d/60~kpc)}^{-1}$. The distribution of derived cloud pressures is shown in the lower right plot of Figure~\ref{halo_density_fig} (bottom axis), with most (68\%) of the values being $P/k = 10^{1.75-2.2}~K $cm$^{-3}$ and the median value begin $P/k = 10^{1.9} K $cm$^{-3} ~\rm{(d/60~kpc)}^{-1}$.

\section{Derivation of Physical Properties: Tests with Simulations}\label{simulation}
In this section we use a set of HVC simulations \citep[see][]{Heitsch09} to assess our choice of temperature (\S 5.1) and our derivations of cloud volume densities (\S 5.2) and pressures (\S 5.3).   We examine the variation in these properties for a range of cloud models and viewing angles, i.e. the angle between cloud trajectory and the line of sight. 

The details of the three-dimensional HVC simulations are published in \citet{Heitsch09}, and the reader is referred to this paper for the details.  The names of the HVC models in Figures ~\ref{f:modeldv} -~\ref{f:modelprs}  and~\ref{f:hdsim} represent the various conditions tested for HVCs and the type of simulation.
The beginning letter of W represents a simulation where the cloud is subjected to a wind of constant density to simulate the movement of the cloud through a diffuse, hot ($10^6$ K) halo medium, and the beginning letter of H represents a possibly more realistic setup, in which the cloud moves through a range of halo densities.  The simulated cloud was setup to have typical observed cloud properties at a distance of 10 kpc, though we tested a range of halo densities as seen in Figure~\ref{f:hdsim}. 
%Series W corresponds to the classical wind tunnel experiment, in which a cold
%cloud is placed in an elongated box on a gravitational potential slide. At $t=0$, a "wind" is being ramped up at the lower
%end of the simulation domain to approximate a cloud moving through a diffuse, hot background medium at observed HVC velocities.
%Series H is possibly a more realistic setup, in which the cloud's progress through the diffuse halo medium is actually followed.
For all simulations, the clouds are gradually disrupted by dynamical instabilities and we examine the clouds when they have developed a mild head-tail structure (i.e. somewhat elongated, but not beyond our aspect ratio of 1.6 limit when examined at high viewing angle values).  The simulation data considered includes all HVC gas with $T<10^4$~K, as a proxy for neutral hydrogen. 

%Why is this relevant for our discussion of cloud pressures and (in the discussion) of halo densities? At issue is here what to chose
%as a "correct" measure for the pressure, and to justify the choice of a constant temperature of $T=9000$~K.

\subsection{Thermal Linewidth Assessment}\label{ss:compdelta}

The narrow distribution of linewidths for HVCs is consistent with a warm neutral hydrogen component with a temperature
of approximately 9000 K.   We begin by assessing how the linewidths may be affected by projection effects and any non-thermal
component by ``observing" several simulated clouds at viewing angles between 0 to 90 degrees between the line-of-sight and the cloud trajectory.   The "observed" linewidth for the simulated clouds can be represented by,
\begin{equation}
  \Delta_{obs} = \sqrt{\Delta_{nt}^2+\Delta_{th}^2},
  \label{e:deltaobs}
\end{equation}
consisting of a non-thermal and thermal component. 

Figure~\ref{f:modeldv} demonstrates the effect of a variation in viewing angle on $\Delta_{obs}$ compared to the thermal linewidth,  $\Delta_{th}$.   The size of the symbols denotes the angle between the cloud trajectory
and the line of sight for $90$ (largest symbol), $60$, $30$ and $0$ degrees.  In other words, the smallest symbols stand for 
clouds moving directly along the line of sight.  The effect of an increase in viewing angle is the same in all models.  The non-thermal component, or largely the disruption of the cloud in the form of a decelerated tail, increases as the viewing angle decreases.   At large viewing angles, the linewidth is a close approximation of the thermal linewidth, while at small viewing angles the observed linewidth can differ by more than a factor of two in the extreme cases where the tail is a large fraction of the cloud's total mass.   This comparison shows that the thermal linewidth is within a factor of 1.5 of the observed linewidth for over 75\% of the clouds.  Going back to our observations, with our adopted cloud temperature of 9000 K, we expect a linewidth of $\sim$20 \kms~ if it is entirely thermal.   Since the vast majority of the GALFA - \HI~clouds have linewidths between 20 - 30 \kms, the simulations indicate this temperature is consistent with the observed linewidth.

\subsection{Dependence of Volume Density on Viewing Angle}\label{ss:compdens}
Our derived volume densities can also be assessed by examining the simulation data at various viewing angles.   
Since as discussed we have set the temperature at 9000 K, the
derived volume density is the only thing that causes a variation in our derived cloud pressure.
 The mean "true" volume density for each simulated cloud ($n_c$)
is easily determined by averaging over the volume density of all gas with $T<10^4$~K as a proxy for neutral hydrogen.
To determine the "observed" volume density ($n_l$), we project each cloud for the angles $0$, $30$, $60$ and $90$ degrees (as above),
and determine the cloud mass by summing up the column densities of each resolution element. This mass $M_c$ is then used to 
determine the volume density, via
\begin{equation}
  n_l = \frac{3 M_c}{4\pi R^3},
\end{equation}
where we use 
\begin{equation}
  R=\sqrt{A/\pi}.
\end{equation}
The cloud area $A$ is determined from the projection and we assume that our cloud is roughly spherical as in the derivation for the GALFA - \HI~clouds. As discussed above,
the accuracy of the resulting density $n_l$ will thus depend on the elongation (or aspect ratio) of the cloud.  Figure~\ref{f:modelnl}
shows the resulting $n_l$ against the actual mean cloud density $n_c$ as determined directly from the 3D density field. Red symbols
denote aspect ratios $<1.6$, i.e. the same selection criterion as used for the observed clouds. Most $n_l$ with this selection
reproduce the "true" mean density within a factor of $3$. 
%Larger aspect ratios lead to a systematic underestimate of the volume density. 

\subsection{Assessing Cloud Pressures}

We use the adopted temperature and volume density to derive cloud pressures, so in this section we use the simulations to compare the actual cloud pressures to deriving the cloud pressures in the same way we do for the GALFA - \HI~clouds.   As noted in the previous two sections, viewing angle and  variation in the cloud temperature are the two main factors that could lead to these values being different.

Figure~\ref{f:modelprs} compares the actual thermal pressure within the three-dimensional simulated cloud to two other pressure estimates.    The top panel shows actual thermal pressure, $P_{th}$, against the derived pressure using the temperature of $T=9000$~K and the "observed"
volume density ($n_l$ from above), or
\begin{equation}
  P_9 = k_B\,9000\,n_l
\end{equation}
This plot shows that {\em underestimates} of the actual pressure (anything below the solid line) are due to large aspect ratios and subsequently too low of volume density (see Fig.~\ref{f:modelnl}).   Since we do not consider these clouds with large aspect ratios, this suggests we are unlikely to underestimate the pressure for the GALFA - \HI~clouds.  The red symbols denoting aspect 
ratios $<1.6$ are in most cases {\em above} the solid line, indicating overestimates. Since almost all of the $n_l$ values are underestimates, {\em overestimates} must be caused by the temperature
of $9000$~K being too high. This is not surprising, given the lower thermal linewidth shown in Figure~\ref{f:modeldv} and the fact that the
model clouds consistently show a two-phase medium that is not evident in this population of GALFA - \HI~clouds (see \S 6.3).   Therefore the overestimates in $P_9$ compared to $P_{th}$ may be partially due to the simulated clouds having a lower average temperature (typically 6600-6700 K) than the GALFA - \HI~clouds.

The bottom panel shows the above $P_9$ estimate of the thermal pressure combined with the non-thermal component derived from the 
non-thermal linewidth,
\begin{equation}
  P_{9t} = P_9 + 3n_l\,\Delta_{nt}^2.
\end{equation}
The estimates are consistently higher as expected with the added non-thermal component, largely caused by the disrupted tail of the simulated cloud.   Given, the non-thermal component is due
to this decelerated tail, it cannot really be interpreted as a "turbulent pressure" and $P_9$ is the more accurate representation of the actual cloud pressure.   The bottom panel also shows the spread between the minimum and maximum angle of one model is now larger than for $P_9$, since the lag between head and tail gets
more dominant in velocity space with smaller angles.

\section{Discussion}

\subsection{Comparison of the Complex C and Magellanic Stream Cloud Populations}

Since we know that the Magellanic Stream (MS) is gas stripped from the Magellanic Clouds \citep{Mathewson74}, by comparing the properties of the Complex C and Magellanic Stream clouds we may gain insight into the origin of Complex C. 
Similar properties, such as density, mass, size, and linewidth distributions, may indicate a common physical process played a role in their origin.  We compare the Stream clouds to both the original resolution Complex C cloud catalog and the smoothed Complex C cloud catalog.  The smoothed catalog may be a better match to the clouds being resolved at the distance of the Magellanic Stream.

\begin{figure}
	\begin{center}
	\includegraphics[scale=1.2]{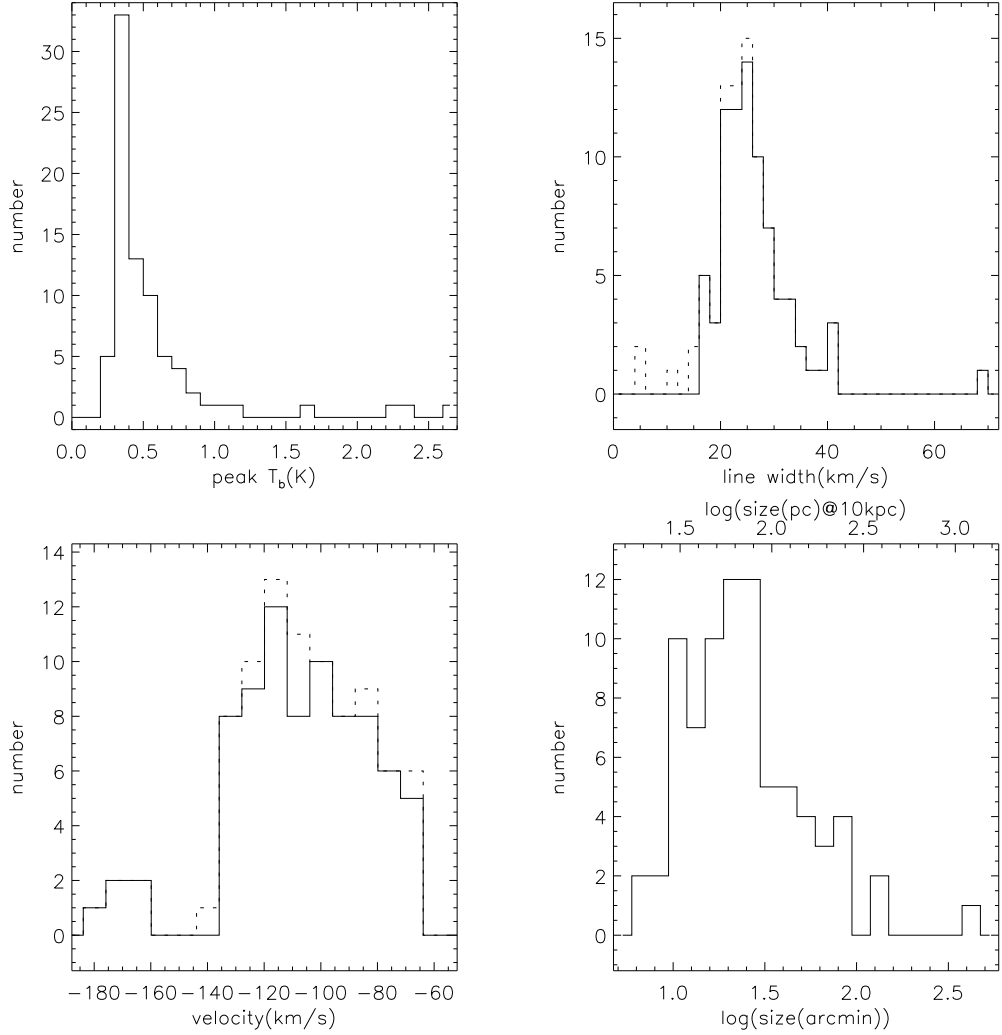}
	\caption{Statistics of the clouds in the Complex C cube (only the clouds with \vlsr \ $<$ -65\kms\ are included). The plots are histograms of peak T$_b$ (K), line width, central velocity (LSR), and angular size and corresponding physical size at 10 kpc. The solid lines in the histograms represent the distribution of clouds that were fit with one Gaussian and the primary components of the clouds that were fit by two Gaussians (the component that contains more mass); the dashed lines include both components of the clouds fit with two Gaussians.}
	\label{C_results1}
	\end{center}
\end{figure}

\begin{figure}
	\begin{center}
	\includegraphics[scale=1.2]{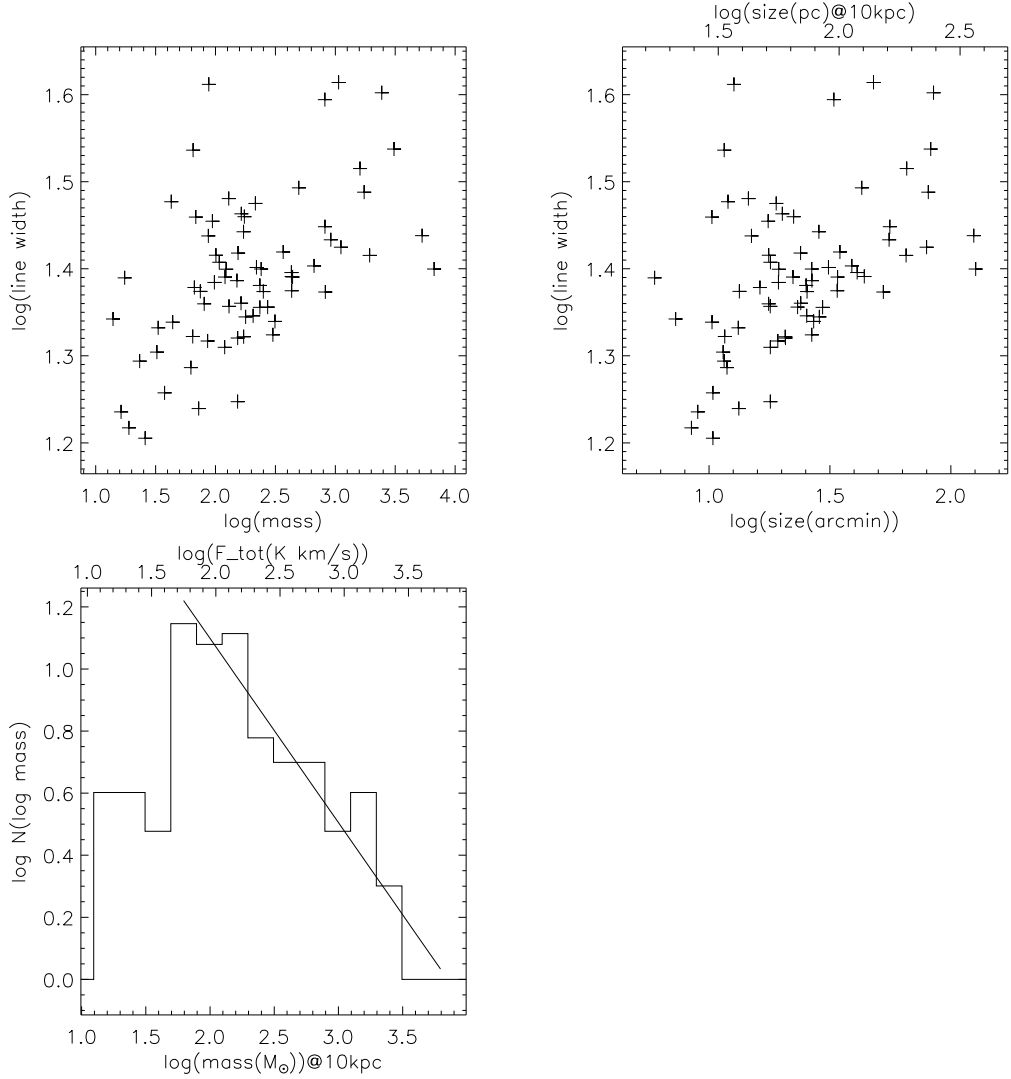}
	\caption{Statistics of the clouds in the Complex C cube (only the clouds with \vlsr \ $<$ -65\kms\ are included). The upper left plot shows line width vs. mass, and the upper right plot is line width vs. size. The clouds with two Gaussian components are removed from the plots for the line width vs. mass and the line width vs. size plots since no single line width can be defined. The lower left is a mass histogram at 10 kpc (slope = -0.60 $\pm$ 0.05 $\log(\rm{N}(\log(\rm{mass})))/\log(\rm{mass})$). }
	\label{C_results2}
	\end{center}
\end{figure}

\begin{figure}
	\begin{center}
	\includegraphics[scale=1.2]{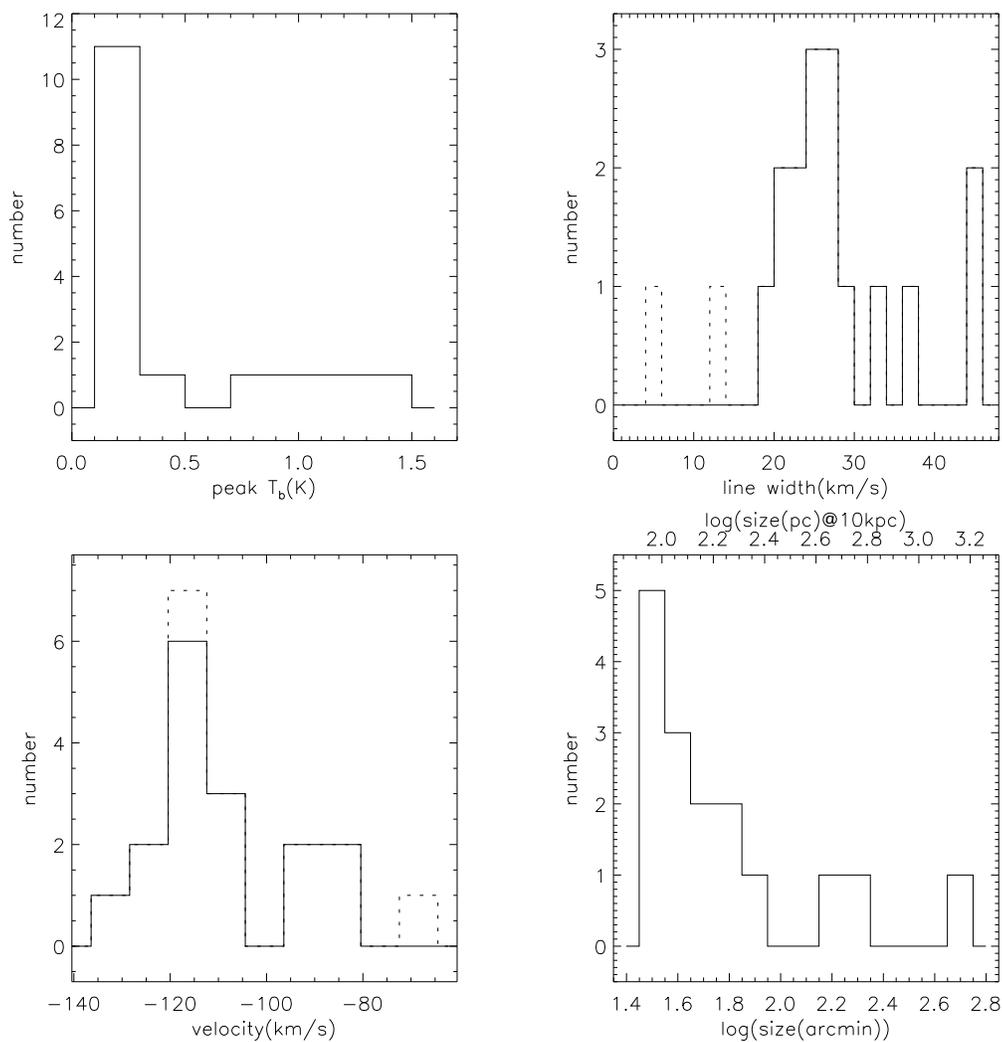}
	\caption{Statistics of clouds cataloged in the smoothed Complex C cube.  The plots show histograms of peak T$_b$ (K), line width, central velocity (LSR), and size. The solid lines in the histograms represent the distribution of clouds that were fit with one Gaussian and the primary components of the clouds that were fit by two Gaussians (the component that contains more mass); the dashed lines include both components of the clouds fit with two Gaussians.}
	\label{CCS_results1}
	\end{center}
\end{figure}

\begin{figure}
	\begin{center}
	\includegraphics[scale=1.2]{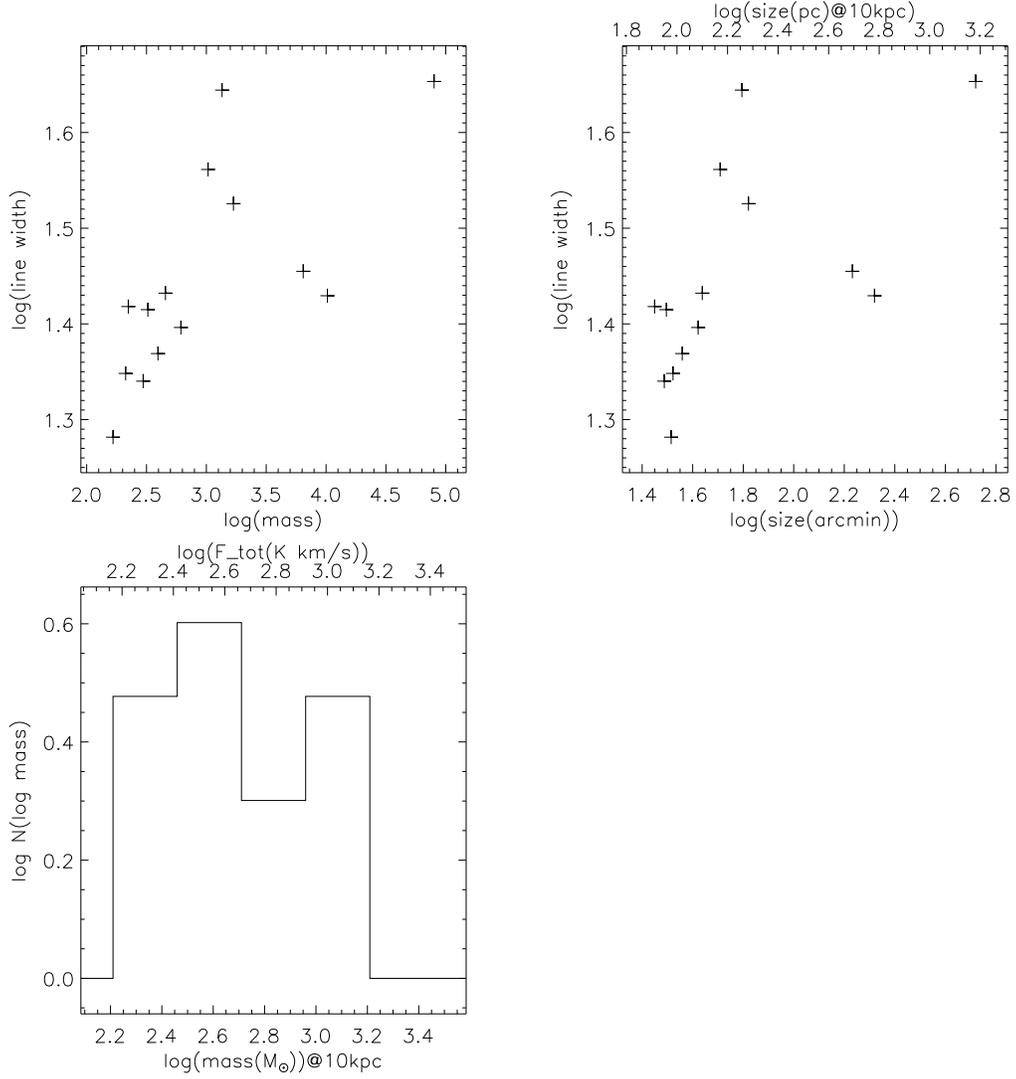}
	\caption{Statistics of smoothed Complex C clouds with \vlsr~$<$ -65\kms\ following the same format as Figure 8.  The upper left and right plots show line width vs. mass and size, respectively.   The bottom figure shows the distribution of masses at 10 kpc. }
	\label{CCS_results2}
	\end{center}
\end{figure}

\begin{figure}
	\begin{center}
	\includegraphics[scale=1.2]{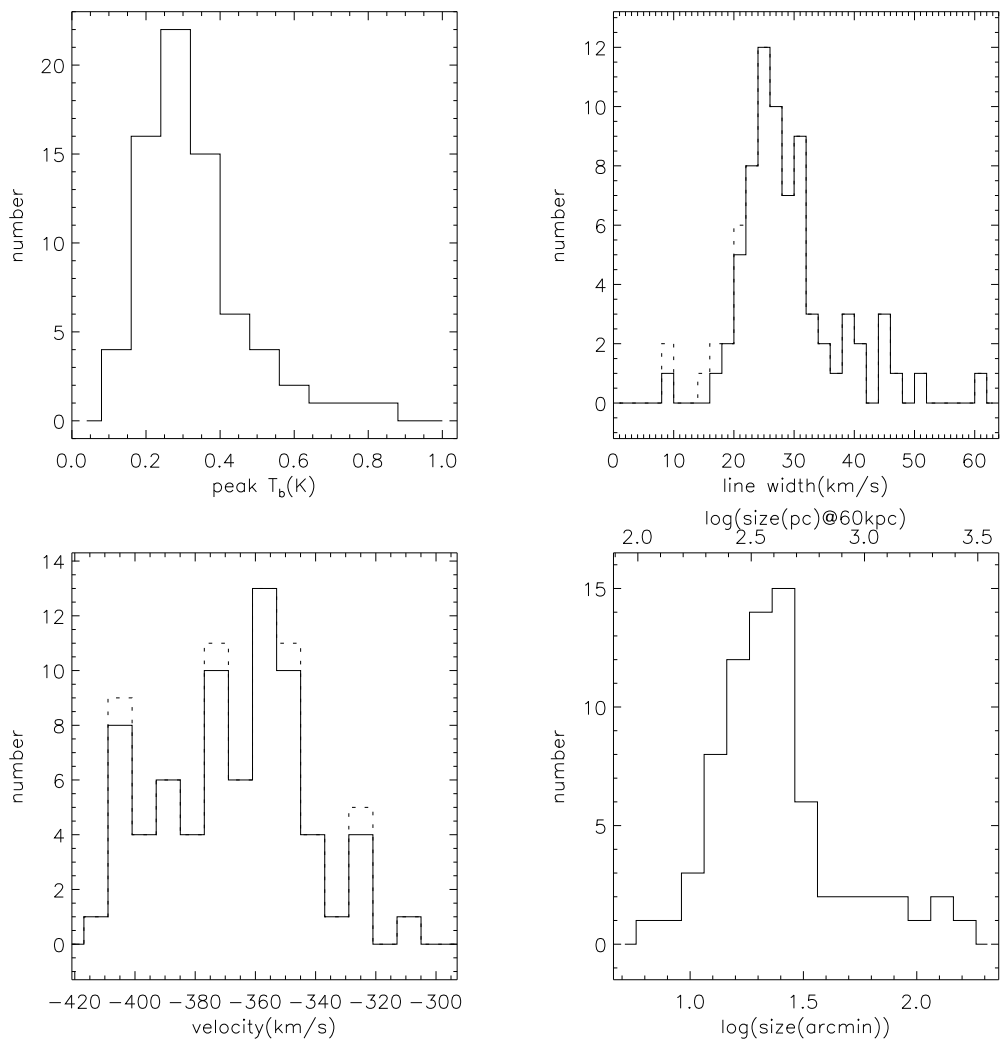}
	\caption{Statistics of Magellanic Stream clouds. The plots show histograms of peak T$_b$ (K), line width, central velocity (LSR), and size in both arc minutes and physical size at 60 kpc. The solid lines in the histograms represent the distribution of clouds that were fit with one Gaussian and the primary components of the clouds that were fit by two Gaussians (the component that contains more mass); the dashed lines include both components of the clouds fit with two Gaussians.}
	\label{MS_results1}
	\end{center}
\end{figure}

\begin{figure}
	\begin{center}
	\includegraphics[scale=1.2]{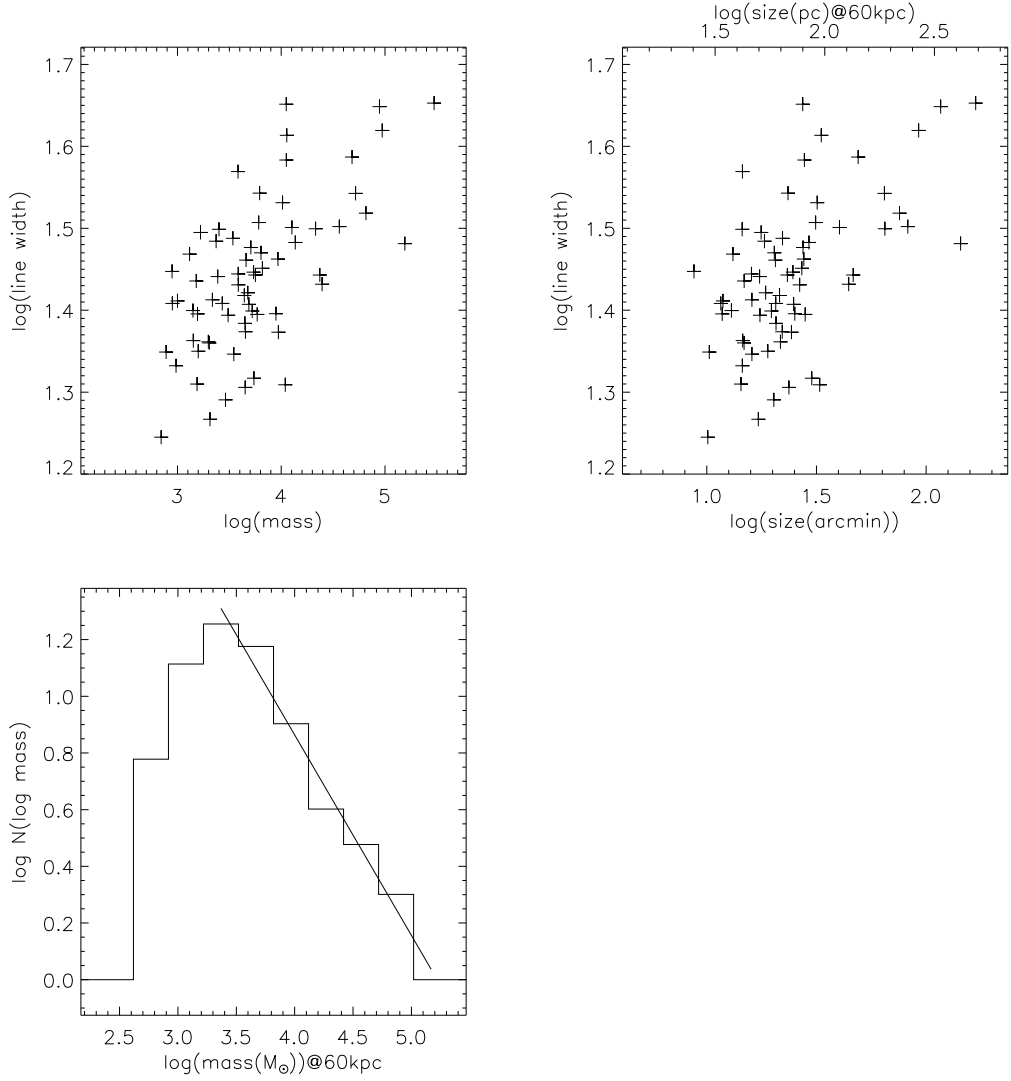}
	\caption{Statistics of Magellanic Stream clouds. The top two plots show line width vs. mass and size. The lower left plot shows mass at 60 kpc (slope =  -0.71 $\pm$ 0.04 $\log(\rm{N}(\log(\rm{mass})))/\log(\rm{mass})$). }
	\label{MS_results2}
	\end{center}
\end{figure}

\begin{figure}
	\begin{center}
	\includegraphics[scale=1.2]{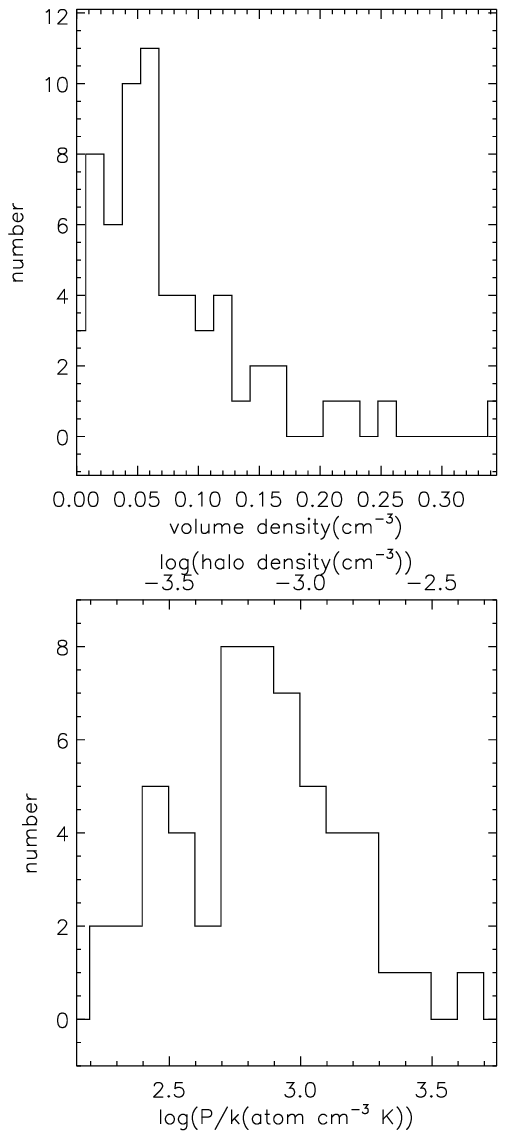}
	\includegraphics[scale=1.2]{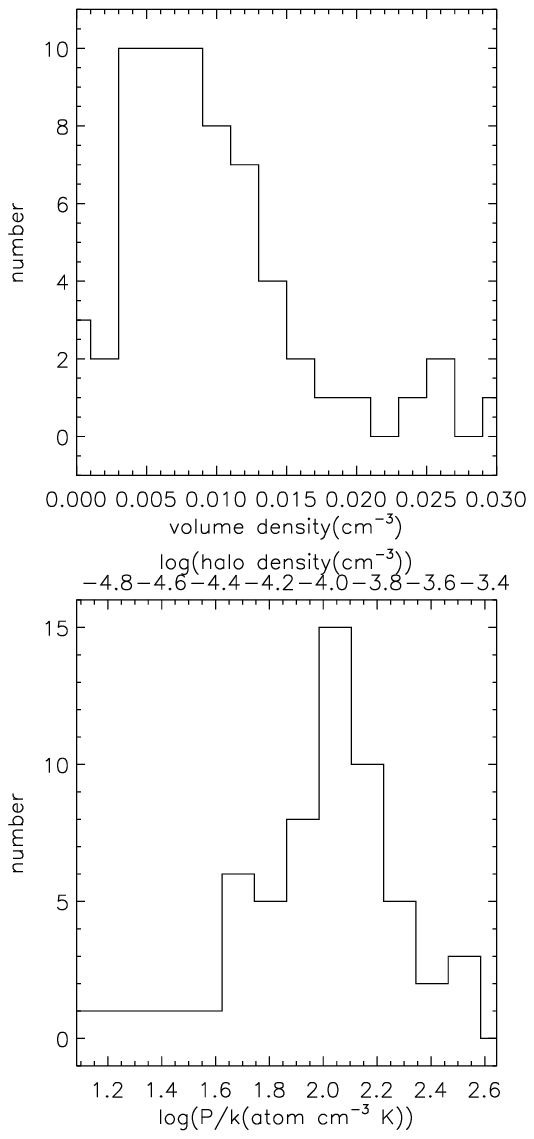}
	\caption{Top: Volume density of the clouds in Complex C (left) and the MS (right) at 10 kpc and 60 kpc, respectively. Bottom: Derived pressure (P/k) and corresponding halo density required to confine the Complex C clouds (left) and MS clouds (right). The temperature of the Galactic halo is assumed to be ${10^{6}}$ K.}
	\label{halo_density_fig}
	\end{center}
\end{figure}

\begin{figure}
  \includegraphics[width=\textwidth]{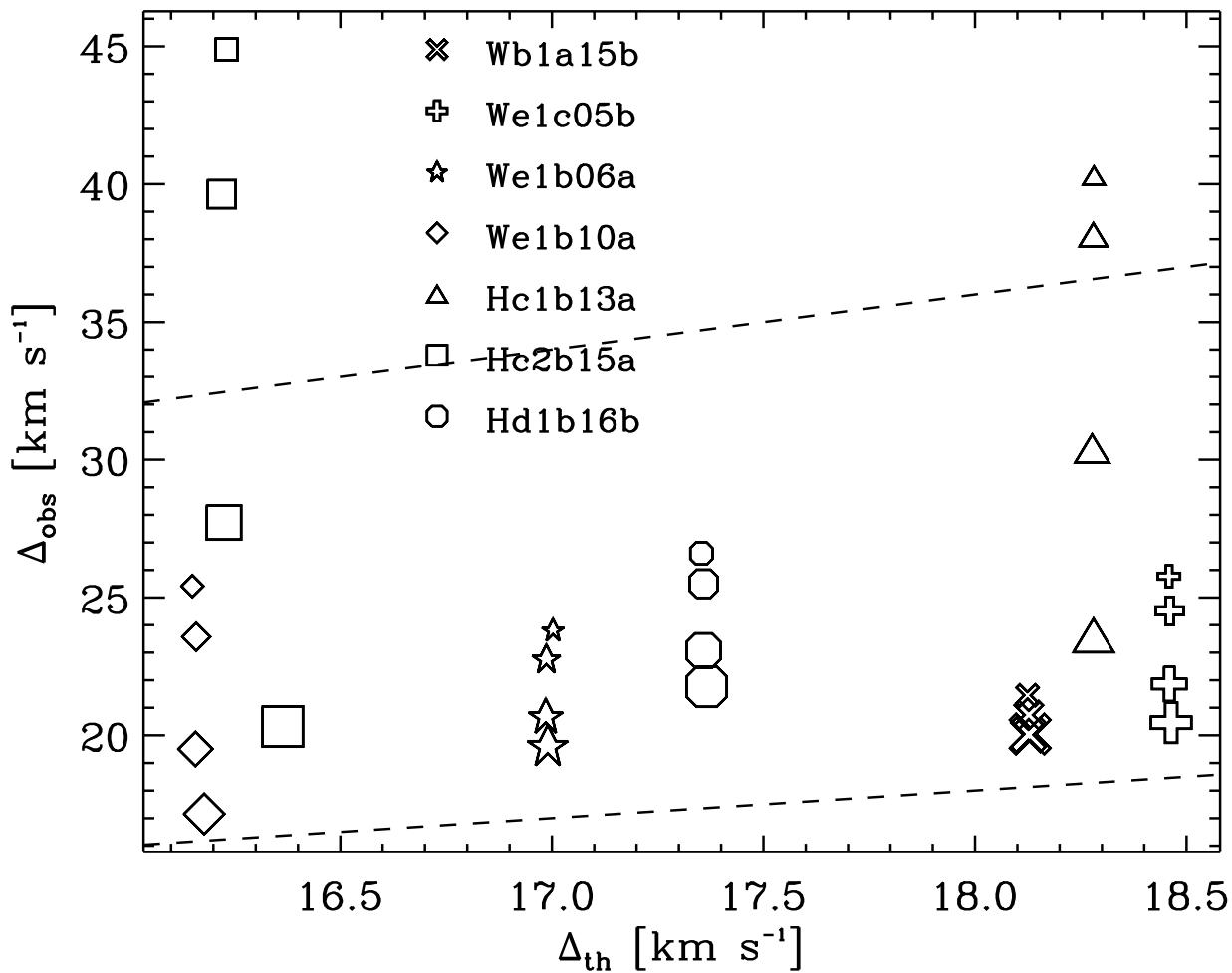}
  \caption{\label{f:modeldv}Total ("observed") line width ($\Delta_{obs}$) against thermal line width ($\Delta_{th}$) for a set of numerical model
           clouds (model names as used by \citealp{Heitsch09}). Symbol sizes denote the angle between cloud trajectory
           and line of sight, with the smallest symbol having the cloud coming at the observer ($0$ degrees) and 
           the largest symbol having the cloud traveling perpendicularly to the line of sight ($90$ degrees).  The dashed lines denote 
           Mach numbers of $1$ and $2$.}
\end{figure}

\begin{figure}
  \includegraphics[width=\textwidth]{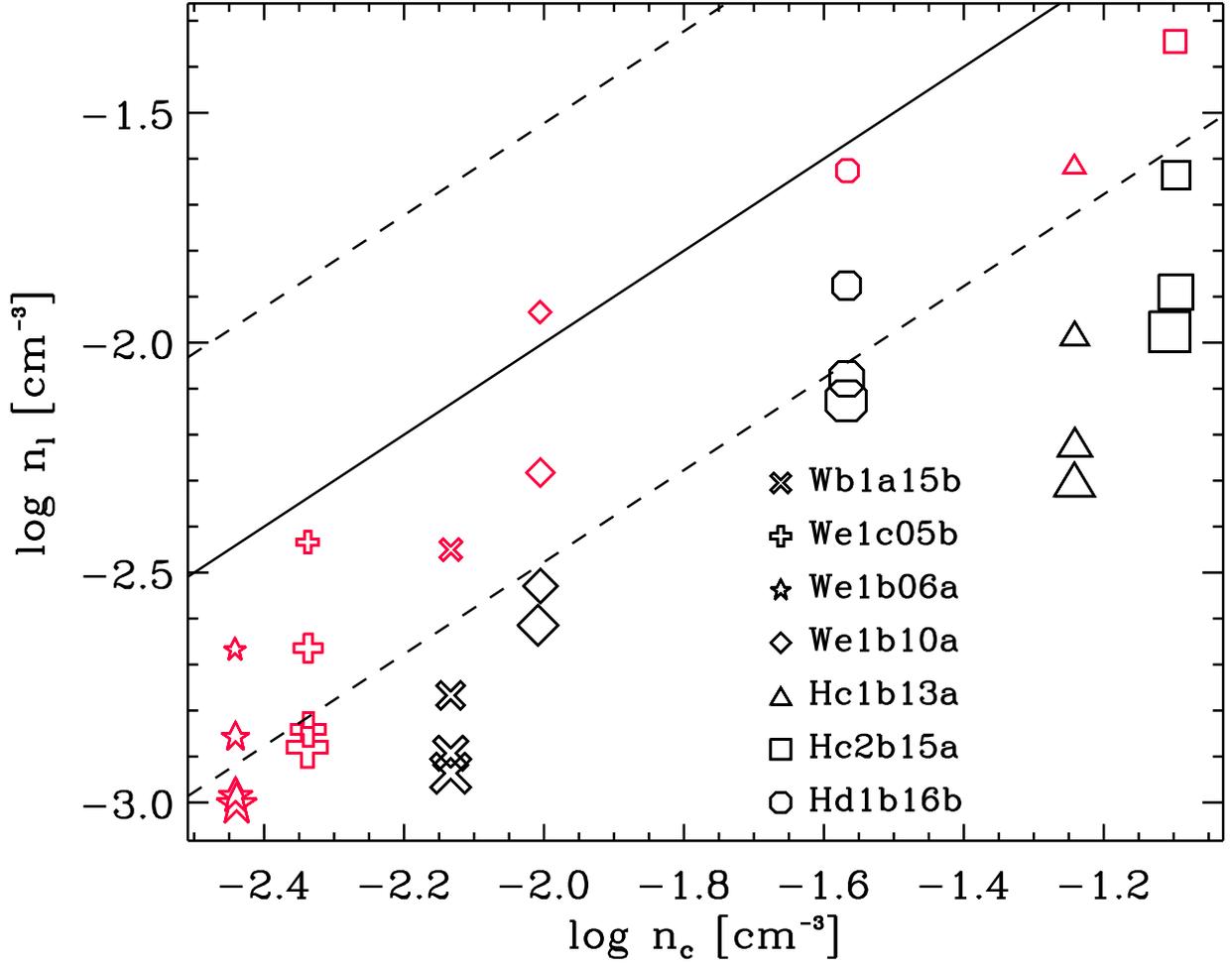}
  \caption{\label{f:modelnl}"Observed" volume density ($n_l$) derived from model cloud column density maps against "true" volume density ($n_c$)
           taken directly from the three-dimensional model clouds. Symbol sizes are as in Figure~\ref{f:modeldv} and red symbols
           denote cloud aspect ratios $<1.6$. The solid line stands for $n_l=n_c$, with the dashed lines showing $3n_c$ and $n_c/3$.} 
\end{figure}

\begin{figure}
  \begin{center}
  \includegraphics[width=0.6\textwidth]{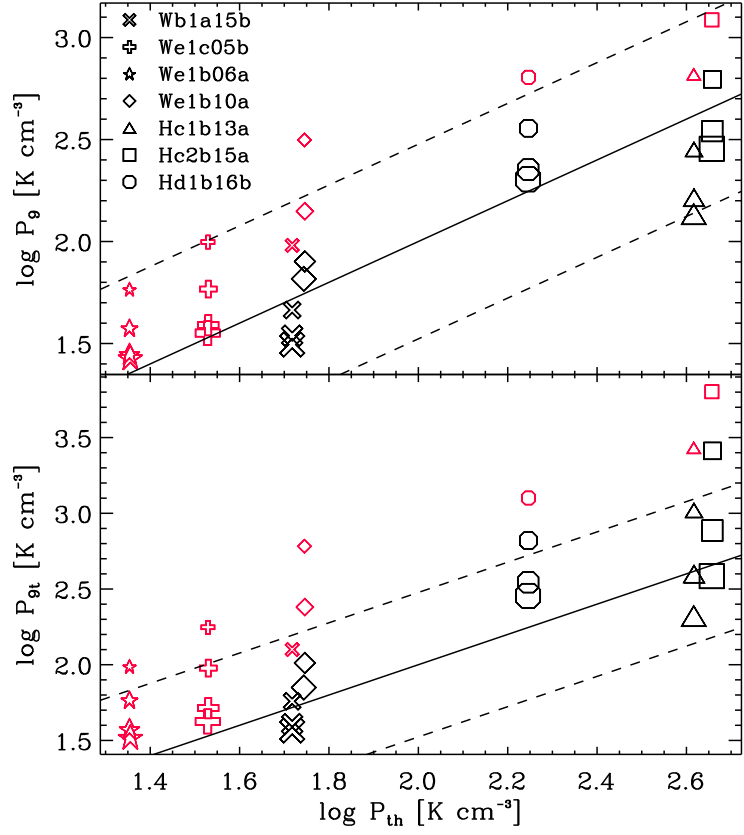}
  \end{center}
  \caption{\label{f:modelprs} Two derivations of simulated cloud pressures, $P_9$ and $P_{9t}$ (see text for a description), against the actual thermal pressure of the cloud, $P_{th}$. Red symbols indicate clouds with an aspect ratio $<1.6$ and symbol sizes are as in Fig.~\ref{f:modeldv}.  The solid line is the two pressures being equal and the dashed lines show $3P_c$ and $P_c/3$.}
\end{figure}

\begin{figure}
  \begin{center}
  \includegraphics[width=0.6\textwidth]{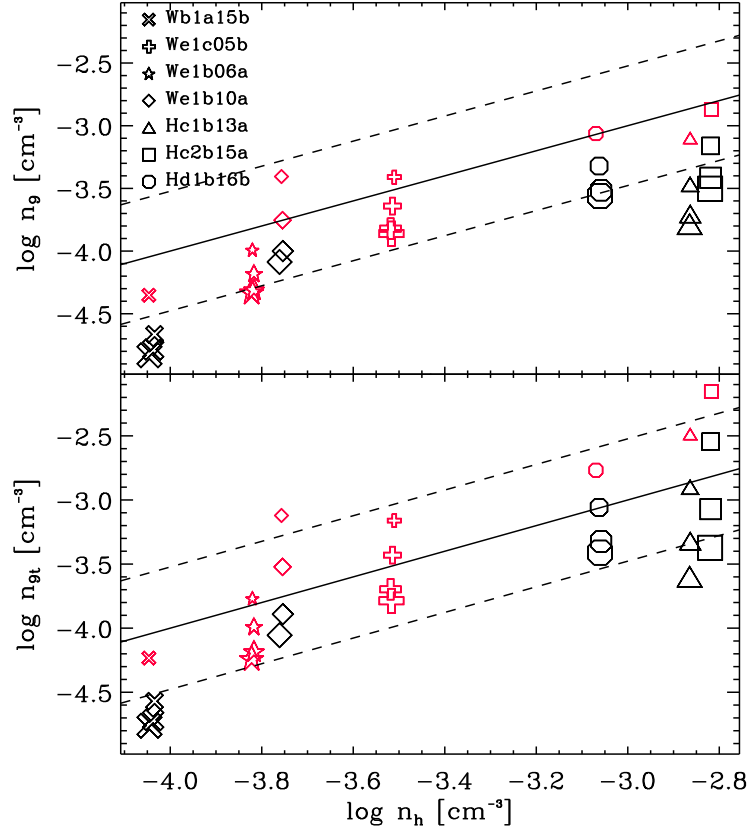}
  \end{center}
  \caption{\label{f:hdsim} Derived halo densities from the cloud simulations, $n_9$ and $n_{9t}$ (see text), compared to the actual halo density in the simulation, $n_h$.  The derived halo densities are based on the pressure estimates in Fig.~\ref{f:modelprs}.  As in previous plots the red symbols indicate clouds with an aspect ratio $<1.6$, symbol sizes are as in Fig.~\ref{f:modeldv}, the solid line is the two halo densities being equal and the dashed lines show $3n_h$ and $n_h/3$.}
\end{figure}

The linewidth distributions (Figures~\ref{C_results1},~\ref{CCS_results1} and ~\ref{MS_results1}) are similar for the MS and Complex C, except for a few outliers that may originate from the blending of two clouds along the line of sight.
The typical linewidth of $\sim$ 20 - 30\kms \ is also found in other lower resolution HVC surveys \citep{DBB02, KH06}, and therefore may indicate a common temperature for not only the MS and Complex C, but the majority of the HVCs.  This linewidth is consistent with a temperature of $\sim$9000 K when the contribution of non-thermal broadening is considered (see \S\ 5.1).  This is also discussed at length in terms of the thermal equilibrium gas temperature for HVCs in W95 and Wolfire et al. (1995a and 2003)\nocite{Wolfire95b}.

The Complex C data (Figure~\ref{C_results1}) and the MS data (Figure~\ref{MS_results1}) have similar angular size distributions in the sense that both complexes show more small clouds than large clouds, with the smallest size limited by angular resolution. Since the physical size of the clouds scales with the distance, the cataloged MS clouds are shifted to larger physical sizes than that of Complex C (100-2900 (d/60~kpc) pc vs. 17-1200 pc).  For the smoothed Complex C cube the size distribution is shifted to 80-1500 pc, with the upper value limited by the area mapped.   

In all the linewidth vs. mass and linewidth vs. size plots (shown in the top panels of Figures~\ref{C_results2}, \ref{CCS_results2} and \ref{MS_results2}) the scatter ($\sim 0.1$) is large compared to the overall range of the linewidth ($\sim 0.4$). Therefore the linewidth does not show clear correlations with the mass and the size of the clouds, and the slight trends (slope = 0.08 and 0.14 respectively for Complex C and 0.12 and 0.23 for MS) are most likely due to selection effects.
% as It is easier to detect a cloud of a given mass with a smaller linewidth or spread over a smaller area on the sky.
In contrast, molecular clouds generally show clear correlations of linewidth vs. mass and linewidth vs. size. \citet{Larson81} found a slope of 0.2 in linewidth vs. mass and a slope of 0.38 in linewidth vs. size. Recent observations give slopes of 0.4 - 0.5 in linewidth vs. size \citep{Elmegreen04,Heyer07}.  The difference is not surprising given the thermal linewidth is a significant component of the total linewidth of HVCs and the relations in molecular clouds are linked through assuming virialization which is unlikely to apply to HVCs (see Sec~6.4).

The \HI \ mass distribution is very similar for the original Complex C clouds and the MS clouds; they both show a power law distribution with a slope of -0.60 and -0.71 respectively. Though the MS cloud distribution is slightly steeper, the slopes are compatible within the statistical uncertainty.   The total intensity range of the clouds in both catalogs is also very similar, although since the complexes are at different distances and the data have different sensitivities the range of masses is different.
%The corresponding mass distributions follow the same power-law, but since the MS is at a farther distance, the corresponding mass range is different. 
If the tail of the MS is at 60 kpc, the mass range probed in the MS is about 20 times higher than the mass range probed in Complex C at 10 kpc ($10^{2.4-5.5}$ vs. $10^{1.1-4.8}$ \Msun).   
%MS data is slightly deeper than CXC data.

%BELONGS IN NEXT SECTION?? The slope predicted by shell fragmentation simulations is -0.4 in log space \citep{WP01}, which is shallower than what we observed.

The median volume density of the MS clouds is 7 times smaller than that of the Complex C clouds (top panels of Figure~\ref{halo_density_fig}, again assuming Complex C is at 10 kpc and the MS is at 60 kpc). Since both the Complex C and MS clouds have a similar range of linewidths, the difference in their densities may be reflected in the cloud pressures.  The median pressure (P/k) derived for Complex C clouds is $\sim 10^{2.8}~\rm{K~cm^{-3}}$, which is about six times greater than that of the median MS cloud at $\sim 10^{1.9}~\rm{K~cm^{-3}}$ (see bottom panels of Figure~\ref{halo_density_fig}).   The pressures derived are consistent with the work of Wolfire et al. (1995a).  The calculations of volume density and pressure are assessed in \S\ 5 through comparison with simulations and it is found that the effect of the viewing angle and the pressure assumptions can cause these values to vary by up to a factor of 3.  This is expected to dominate over other uncertainties given the 13 clouds in the smoothed Complex C cube (showing the effect of a bigger beam and higher sensitivity) results in volume densities less than a factor of 3 lower.   Given our large sample of clouds for original Complex C cube and the MS, we do not expect the entire population to be affected by viewing angle or beam dilution and the distributions will remain distinct.

We note here that the distance to the tail of the Magellanic Stream remains uncertain and this will scale the MS values accordingly.  For simplicity, we have put the tail at roughly the distance of the Magellanic Clouds, though many models have the tail at larger distances.  The tidal stripping models often put the tail of the Stream at distances of approximately 100-200 kpc (e.g. Conners et al. 2006 \nocite{Connors06})
%while ram pressure stripping models claim the tail of the Stream is closer than the Clouds \citep[up to 20 kpc;][]{MD94,mastropietro05}.  The 
and models based on recent proper motion calculations indicate the Magellanic Clouds may be on their first passage and that the tail of the Stream is at $\sim 120$ kpc \citep{Besla07}.
%; although see also Shattow \& Loeb (2009\nocite{SL09}) for a bound orbit with the new proper motions.  
If the Stream is as distant as 120 kpc the most massive clouds cataloged here would be at \Mhi~ $=10^{6.1}$ \Msun, or half as massive as the entire Complex C and 1\arcmin~would correspond to a physical size of 35 pc. 
%If the distance to the Stream is 20 kpc, unlikely with the current proper motion estimates, the most massive clouds would be at \Mhi $ = 10^{4.5}$ \Msun, and 1\arcmin~would correspond to physical size of 6 pc.  
The median volume density and pressure would shift to 0.005 $\rm{cm}^{-3}$ and 42 (K $\rm{cm}^{-3}$) at 120 kpc.
% and 0.2 $\rm{cm}^{-3}$ and 600 ($\rm{cm}^{-3}$ K) at 20 kpc. 

Though the absolute values for the sizes and masses of the Magellanic Stream clouds depend on the actual distance (d for the size and d$^2$ for the mass), the slopes of the distributions will not change, and therefore will remain similar to the slopes of the Complex C cloud distributions.   This includes the original and smoothed Complex C cloud distributions, although there are a limited number of clouds in the smoothed cube.  The relatively narrow distribution of linewidths found here suggests the clouds have a common warm temperature component, and their size may be dictated by the surrounding pressure and cloud mass.  Overall, the similarities in the MS and Complex C distributions are intriguing and may suggest a similar formation or destruction mechanism for the two complexes, despite their different locations in the Galaxy's halo.  
The results also suggest that smaller and lower mass clouds will be detected for both complexes as deeper and higher resolution surveys are completed.

\subsection{Comparison of Complex C Clouds to Galactic Disk-Halo Clouds}

In the Complex C cube, clouds can be cataloged from high negative velocities ($\sim-180$ \kms) into Galactic emission where it becomes difficult to distinguish individual clouds.   This may represent the accretion and breaking up of Complex C as it is approaching the disk, or simply an overlap between the population of clouds in Complex C and at the disk-halo interface.  In any case, a comparison between the properties of the clouds associated with HVCs and those that are at the disk-halo interface is an interesting exercise, as it may provide insight into whether the clouds represent similar gas at different stages of the accretion process.

There are several studies of clouds at the disk-halo interface in the literature.  The study of the disk-halo clouds on the positive velocity side of the GALFA-\HI \ Complex C cube will appear in another paper, but a
population of discrete disk-halo clouds in this general region was found and discussed by \citet{Lockman02}.   He found that these clouds follow Galactic rotation and are discrete in position and velocity, with typical sizes of a few tens of pc. \citet{Stil06} and \citet{Ford08} derived the statistical properties of additional clouds that are most likely at the disk-halo interface using the VLA Galactic Plane Survey and the Parkes Galactic All-Sky Survey. The tangent point method was used to derive the distance to the clouds, and properties such as mass and physical size were subsequently derived. 
The typical size and mass are similar to a smaller Complex C cloud, with the mass distribution peaking around a median mass of 600-700 \Msun.  The mass distribution does not show the power-law distribution we see for both Complex C and the Magellanic Stream, but this may be due to the selection of the disk-halo clouds as discrete and unrelated spatially, while we are studying all clouds related to a single complex.
\citet{SPH06} observed disk-halo clouds in the outer Galaxy toward the Galactic anticenter with GALFA-\HI. Since these observations are taken with the same instrument as the data used in this work, the angular and velocity resolution are the same, allowing a direct comparison of the cloud properties. These clouds are in general colder than the Complex C and MS clouds, 
%and are even colder than those found by \citet{Lockman02} and \citet{Stil06}, though resolution most likely plays a role in the case of the \citet{Lockman02} clouds.  
with an average kinetic temperature of $\sim$ 470 K.
On average, disk-halo clouds have smaller linewidths ($\Delta \rm{V} \approx 13$\kms \ for a median cloud in \citealt{Ford08}; $\Delta \rm{V} \approx 4$\kms \ for a median cloud in \citealt{SPH06}) compared to the typical HVC value of 25\kms, suggesting a lower temperature. A more complete census of disk-halo clouds is being completed with the GALFA-\HI~Survey, and this will allow for more thorough comparisons of the cloud populations.

\subsection{Clouds with Multi-phase Structure}\label{ss:multiphase}
Multiple components in the line profiles of HVCs are commonly observed (e.g., Haud \& Kalberla 2007), and have been explained as the existence of a two-phase structure from calculations of the thermal equilibrium temperature for neutral hydrogen gas under different conditions (W95).   W95 considered the conditions under which thermally bi-stable structures can exist and discussed the corresponding physical environments. 
 In our Complex C catalog, only 8 clouds (10$\%$) require a two component fit to the line profile. 
Following S08, we consider the clouds in our catalog to have a multi-phase structure only when the absolute difference between the \vlsr~of the two components is smaller than the $\Delta \rm{V}$ of the primary component.  This excludes cataloged clouds with two components due to multiple clouds along the line of sight.  For Complex C all 8 ($10 \%$) of the clouds with a two component fit satisfy this criterion. In the Magellanic Stream data, only 4 of the clouds ($6\%$) require a two component fit, and all of them satisfy the criterion for multi-phase structure. 

Since the number of clouds requiring a multiple Gaussian fit to the line profile is lower than that found in previous studies (e.g., S08), we also looked at the line profile at the peak pixel of the cloud.  We were originally considering the integrated spectrum over the whole cloud, which might wash out the narrow-component of the cloud due to velocity structure across the cloud.
With the line profile fit to the cloud's peak pixel, the percentage of clouds in the Stream catalog that require two-component fitting is similar to that found by S08 ($15\%$). This serves as a reality check, showing that the fitting criteria are similar. In the case of the Complex C cloud, when only the peak pixel is considered the same number of clouds require two-component fitting, or $7\%$ of the total. This is consistent with the findings of \citet{KH06} using the LAB Survey with a factor of 10 larger beam than our survey.  They found that about $7.5\%$ of the observed positions have multi-component line profile structure in the negative velocity part of the MS, and only 6$\%$ of the positions in Complex C. It should be noted that we only consider the spectra at the peak pixel of each cloud and \citet{KH06} considered the spectra of all observed positions. 

W95 find in general that in a higher density halo medium a two component profile, or core-halo structure, is expected for the HVCs. It is therefore counter-intuitive that the multi-phase structure is not more prominent in Complex C given that it is closer to the plane of the Galaxy. The slightly lower sensitivity of the Complex C cube compared to the Stream regions is unlikely to account for this difference, as we also fit the peak pixel of the smoothed Complex C cube, which has a higher sensitivity, and there are still few clouds with two-phase structure.  In contrast to Complex C, at the z height of the Magellanic Stream, a two-phase structure is not expected, as largely observed.

One possible explanation is the role of dust content and metallicity in determining the cooling curve (W95).  Complex C has a metallicity range between 0.1-0.5 solar across the cloud \citep{Gibson01, Collins03, Tripp03, Collins07}, but seems to have little or no dust \citep{Ritcher01, Peek08b}.  For, reference, the gas associated with the Magellanic System has a metallicity of about 0.2-0.4 and appears to have some dust \citep{Gibson00,Sembach01}. To understand whether two-phase structure should exist in Complex C, we follow the analysis outlined by W95 and Wolfire et al. (2003, hereafter W03) \nocite{Wolfire03} and calculate the thermally stable condition at the tail of Complex C.

%In the Galactic plane, photoelectric heating is the dominant heating process. In Complex C, since the dust-to-gas ratio is low, it is not obvious whether photoelectric heating will still dominate over other heating processes such as EUV and X-ray heating and cosmic ray heating. However, the contribution from cosmic ray is consistently about one order of magnitude smaller than the contribution from X-ray. As a result, only photoelectric heating and X-ray heating will be considered here. 

We adopt the standard photoelectric heating rate given by W03.  The heating from the FUV field can be written as a combination of the radial FUV intensity given by equation (14) in W03 and the vertical variation given by equation (4) in W95, with the out-of-plane FUV intensity 0.6 times that of the midplane (W95).  Since cosmic ray heating is consistently approximately an order of magnitude lower than FUV/X-ray heating it is not considered here.  At low temperatures, the dominant cooling process in the cold phase is radiative line cooling of the $[\rm{C_{II}}]$ 158 $\mu$m fine-structure transition,
%including collisions with neutral hydrogen and electrons.
while at higher temperatures, cooling is the result of several lines: $[\rm{C_{II}}]$ 158 $\mu$m, $[\rm{O_{I}}]$ 163 $\mu$m and Ly$\alpha$.  The cooling rate for each of these species using a metallicity of 0.3 solar is given by W03.   In this simple model of heating and cooling in Complex C, a multi-phase structure is predicted to exist.  At log(P/k) = 3, Complex C has two thermally stable equilibrium temperatures, a cold and warm component.   Observationally, the majority of the clouds in this region of Complex C show only the warm component. 
%The temperature of the warm component is regulated mainly by Ly $\alpha$ emission and is about 9000 K.

The fact that the Complex C clouds show only the warm component may be due to a combination of long cooling times and the dynamic, turbulent environment of the complex.
The dominant coolant $[\rm{C_{II}}]$ has a cooling rate on the order of $2\times10^{-28}$ erg/s around $\rm{n \approx 0.06\ cm^{-3}}$, the median density we found for the Complex C clouds.  The cooling time is then approximately $\rm{tc \approx [(3/2) \ kT]/(2\times10^{-28}~erg/s) \approx}$ 300 Myrs.   This is long compared to the typical lifetimes expected for halo clouds moving through a diffuse halo medium (generally $<$80 Myr; Heitsch \& Putman 2009)\nocite{Heitsch09}.  Therefore the clouds may not live long enough to cool substantially and form a two-phase structure.

\subsection{Inference on Halo Density}\label{halo_density}

HVCs are often assumed to be in pressure equilibrium with the surrounding hot halo medium, and we use that assumption here to infer the density of the diffuse gaseous halo.   The clouds are unlikely to be gravitationally confined, as using $\rm{M_{dyn} \ge R \Delta \rm{V}^{2}/8G}$ and the typical properties of Complex C clouds (R = ${10^{1.7}}$ pc, $\Delta \rm{V} = 25$\kms), we find a mass 1000 times greater than \Mhi~($\sim{10^{2.2}}$ \Msun).  This is also consistent with the lack of linewidth-mass correspondence (Figures~\ref{C_results2}, \ref{CCS_results2} and \ref{MS_results2}).
Assuming pressure balance between the cloud and the confining halo medium, the external pressure is given by ${P_{h}=kn_{h}T_{h}}$, where ${T_{h}}$ is the halo temperature and is assumed to be ${10^{6}}$ K, and ${n_{h}}$ is the halo density.  As outlined in \S~\ref{results}, the thermal pressure of the cloud is given by ${P_{c}=kn_{c}T_{c}}$, where ${T_{c}}$ is the cloud temperature set at 9000 K, and ${n_{c}}$ is the cloud volume density for clouds with an aspect ratio less than 1.6. 

Figure~\ref{halo_density_fig} shows the pressure of the clouds in units of (P/k) and the halo density required to confine each cloud if they are in pressure equilibrium.  
The required halo densities lie within a reasonable range.  For Complex C at a distance of 10 kpc or $\sim$8 kpc from the Galactic Center and a z-height of 2 - 4 kpc, P$/$k is typically $10^{2.6}$ to $10^{3.1}$ K cm$^{-3}$.  This corresponds to a halo density range of ${10^{-3.3}}$ to ${10^{-3}\ \rm{cm}^{-3}}$ at a temperature of $10^6$ K. The scatter in the inferred halo density most likely shows the uncertainty in the spherical symmetry and pressure equilibrium assumptions, rather than indicating a large variation in halo density.  
% See Sec.~\ref{simulation} for details.
%Need quantitative estimate of the scatter.

We can do the same analysis for the Magellanic Stream, with the consideration that the distance is more uncertain for the tail of this complex.  At the assumed distance of 60 kpc, or a z height of $\sim$40 kpc, P$/$k is typically $10^{1.75}$ to $10^{2.2}$ K cm$^{-3}$, corresponding to a typical halo density of $10^{-4.1}$ to $10^{-3.7}~\rm{cm}^{-3}$. This value is consistent with the results of previous studies of the Magellanic Stream \citep{SDK02}.  Note that if we change the distance by a factor of 2, P$/$k and halo density would change by a factor of 2 as well.   Therefore if the tail of the Stream is actually at 120 kpc, the surrounding confining halo density would need to be only  $10^{-4.4}$ to $10^{-4.0}~\rm{cm}^{-3}$.

In \S~5 we assess the use of our derived volume density and temperature value to obtain a cloud pressure through an investigation of cloud simulations ``observed" from various viewing angles.  We can now compare the actual halo densities in the simulations to values obtained by setting the thermal pressure of the surrounding halo gas to the observed cloud pressures ($P_9$ and $P_{9t}$ from \S~5.3).
The results of the comparisons are shown in Figure~\ref{f:hdsim}.  The top panel shows the halo densities ($n_9$) derived using the $P_9$ cloud pressures (similar to the pressures derived for the GALFA - \HI clouds) vs. the actual halo density ($n_h$).  The clouds with the smallest aspect ratios give the best estimates, though all of the clouds that fit into our aspect ratio cut are within a factor of 3 of the actual halo density. In the top panel the derived halo density is almost always lower than the actual halo density.   The bottom panel shows the halo densities ($n_{9t}$) obtained using $P_{9t}$ values, or $P_9$ plus a possible pressure contribution from the non-thermal linewidth, against $n_h$.
The clouds with the smallest aspect ratios shift to being overestimates of the halo density, consistent with the non-thermal linewidth not representing a form of pressure that should be included in the analysis.   As with $P_{9t}$, the scatter also increases for $n_{9t}$ compared to $n_9$.

Our derived halo density estimates can be compared to values obtained from other observations and from simulations.  Observationally, estimates of halo density have been derived from pulsar dispersion measures, O VI and O VII observations, and HVC structures. The pulsar dispersion measures are thought to be a direct way to detect the halo medium. \citet{Gaensler08} derived the 3-$\sigma$ upper limit of the halo density to be $10^{-3.1} \rm~{cm}^{-3}$ from pulsar dispersion measures at $z > 5$ kpc. They also derived the distribution of the WIM to be $n(z) = 0.031 \exp(-z/1~\rm{kpc})\ \rm{cm^{-3}}$, which gives a density of $10^{-2.8}~\rm{cm^{-3}}$ at $z \sim 3$ kpc. This is consistent with the halo density we derive. 
%\citet{Peek07} derived a halo model from the structure of the HVCs, which gives a density of $\sim 3 \times 10^{-4} \rm cm^{-3}$ at 10kpc. 
At the distance of the Stream, the halo density we derive is consistent with O VI and O VII observations \citep{Bregman07,Sembach03} and the densities needed to strip gas from dwarf galaxies \citep{Grcevich09}.

On the theoretical side, models of the distribution of hot diffuse halo gas are generally spherical. Both $\Lambda$CDM cosmological simulations and analytical approximations of a Milky Way-like galaxy predict a halo density of  $\sim 10^{-3} - 10^{-3.8}~\rm cm^{-3}$ at 10 kpc \citep{MB04, Sommer-Larsen06, Kaufmann08}. These values are roughly consistent with our halo density estimates from the Complex C observations at $\sim$ 8 kpc from the Galactic center, though these simulations are not designed to accurately represent the region near the disk. 
At 50 kpc, the predicted density from the simulations is about $\sim 10^{-4.2} - 10^{-3.5}~\rm cm^{-3}$, which is similar to our estimate from the Magellanic Stream.  It should be noted that if the halo temperature is actually closer to $2\times10^6$ our halo densities will decrease by a factor of 2.  This is likely to be a larger effect than a small variation in the temperature of the warm halo clouds.

\section{Conclusions}

This paper presents new \HI \ observations of the tail of Complex C from the GALFA-\HI \ Survey. The observations have 3.5\arcmin~spatial resolution and 0.18\kms \ channel spacing, smoothed to 1.4 \kms \ for this work. We catalog discrete clouds at the tip of Complex C and the Magellanic Stream with Duchamp, an automated source finder that provides a systematic and objective way of cataloging clouds.  Diffuse Galactic emission was removed from the Complex C data before cataloging to enable us to search for clouds at lower velocity and obtain accurate \HI \ column densities for the clouds.  The cloud spectra are fit with Gaussian profiles to obtain their LSR velocities, FWHM values, and column densities and, given we have a distance for Complex C and a distance estimate for the Magellanic Stream, their physical properties are derived, including mass, size in parsecs, volume density, and pressure.  We then compare the physical properties of the clouds in the Magellanic Stream and Complex C (including the original cube and a smoothed cube to approximate Complex C at a distance more appropriate to the Magellanic Stream), and derive the density of the surrounding halo medium assuming pressure balance.  Finally we test the derived properties with three dimensional simulations of clouds moving through a diffuse medium.  Several of the main results can be summarized as follows: 
\begin{itemize}
\item{The Complex C and Magellanic Stream clouds show similarities in their linewidth, angular size, and mass distributions.  The common linewidth of $\sim25$\kms \ is found for HVCs in general and is indicative of a warm component for the clouds.  Both complexes show a power-law distribution of mass above the detection limit. The Complex C clouds have a mass distribution of slope = -0.60 $\pm$ 0.05 and the Magellanic Stream clouds have a slope = -0.71 $\pm$ 0.04. The clouds in the two complexes do not show any distinctive differences and the similarities suggest a similar origin of formation or common physical process breaking down the clouds.  This is despite their different halo environments in terms of distance and the derived surrounding pressure.}
\item{The Complex C clouds have a median $T_{b, peak}$ of 0.41 K, median linewidth of 24.9 \kms, median \vlsr of -105 \kms and median size of 22.2 \arcmin, corresponding to 64.4 pc at a distance of 10 kpc. The smoothed Complex C clouds have a median $T_{b, peak}$ of 0.18 K, median linewidth of 26.2 \kms, median \vlsr of -113 \kms and median size of 51.7 \arcmin, corresponding to 150 pc at  a distance of 10 kpc. The MS clouds have a median $T_{b, peak}$ of 0.30 K, median linewidth of 27.7 \kms, median \vlsr of -364 \kms and median size of 22.1 \arcmin, corresponding to 376 (d / 60 kpc) pc.}
\item{From Gaussian fitting of the line profiles, it is found that neither the Magellanic Stream or Complex C clouds have abundant two-phase structure (7 \% of the clouds).  Following the analysis of W03 and W95 we confirm that a two-phase structure is expected in the environment of Complex C.  The lack of this structure may be explained by the fact that Complex C has a low metallicity and thus the cooling time is long compared to a typical cloud's lifetime.}
\item{Assuming the clouds are confined by the pressure of the surrounding hot halo medium, we estimate the density of the hot halo medium at the z-height of the tail of Complex C ($\sim3$ kpc) is $10^{-3.3}- 10^{-3.0}~\rm{cm^{-3}}$, with a median value of $5.8~\times~10^{-4}~=~10^{-3.2}~\rm{cm^{-3}}$.  For the Magellanic Stream we obtain values of $10^{-4.1} - 10^{-3.7}~\rm{cm^{-3}}$ (with a median value of $10^{-4.1}~\rm{cm^{-3}}$) at a z-height of 40 kpc, and this would scale down by a factor of two if the distance is a factor of two greater.  These estimates are consistent with previous observations and models of the Galactic halo.}
\item{We assess the derived physical properties or our clouds with simulations. We justify the use of a constant temperature of $9000$K, as well as a selection of an aspect ratio of $<1.6$ for deriving the volume densities, pressures, and halo densities from the clouds. The analysis finds that these properties are accurate to within a factor of $3$. }

\end{itemize}

\paragraph{Acknowledgements\\}
We thank Jana Grcevich for help with the observations, Kevin Douglas for help with the data reduction, and the rest of the GALFA-\HI~Survey team for the development of the tools that helped in the collection and reduction of this data.
This research made use of the Duchamp source finder, produced at
the Australia Telescope National Facility, CSIRO, by M. Whiting.
MEP, SS, and JEGP acknowledge support from NSF grants AST-0707597/0917810, 0707679 and 0709347. FH acknowledges support from NSF grant AST-0807305.
MEP also acknowledges support from the Research Corporation.
 We credit the use of the Karma visualization software \citep{Gooch96}. The Arecibo Observatory is part of the National Astronomy and Ionosphere Center, which is operated by Cornell University under a cooperative agreement with the National Science Foundation.

 \bibliography{./ref} 
 
\clearpage
\begin{deluxetable}{cccccc}
  \tabletypesize{\small}
  \tablewidth{0pc}
  \tablecaption{Cloud cataloging parameters}\label{tab1}
  \tablehead{\colhead{HVC complex}& \colhead{$\Delta$x  \tablenotemark{a}} & \colhead{$\Delta$v \tablenotemark{b}} & \colhead{Detection $\sigma$ \tablenotemark{c}} &
                        \colhead{Grow $\sigma$ \tablenotemark{d}} &\colhead{\vlsr~Range}\\
   		 \colhead{}&\colhead{\arcmin}&\colhead{\kms}&\colhead{K}& \colhead{K}&\colhead{\kms} }
  \startdata
  
   Complex C                   & 3   &  1.42   &  0.29   & 0.18    & $-200- -50$\\
   Complex C (smoothed) & 18 &  1.42   &  0.12   & 0.08    & $-200 - -50$ \\
   MS (region 1)              & 3    &   1.42  &  0.27   & 0.17    &$-420 - -280$ \\
   MS (region 2)              & 3    &   1.42  &  0.15   & 0.09    &$-420 - -280$  \\
    \enddata
    \label{tab1}
     \tablenotetext{a} {\scriptsize Angular resolution of the searched cube.}
    \tablenotetext{b} {\scriptsize Velocity resolution of the searched cube.}
  \tablenotetext{c} {\scriptsize Primary detection threshold.}
  \tablenotetext{d} {\scriptsize Grow threshold: pixels above the grow threshold neighboring a detected pixel will be added to the detection.}
\end{deluxetable}

\clearpage
\begin{deluxetable}{rrrrrrrrrrrrr}
\rotate
\tabletypesize{\scriptsize}
\tablewidth{0pc}
\tablecaption{Complex C Cloud Catalog}
\tablehead{\colhead{Cloud \#} &  \colhead{RA} &  \colhead{DEC} & \colhead{l} & \colhead{b} &\colhead{$\Delta RA$\tablenotemark{a}} & \colhead{$\Delta Dec$\tablenotemark{a}} & \colhead{$T_{b,peak}$} &  \colhead{\vlsr} & \colhead{\vgsr}& \colhead{$\Delta V$} &  \colhead{\NH(tot)} & \colhead{\Mhi\tablenotemark{b}} \\
 &\colhead{(J2000)} &    \colhead{(J2000)} &\colhead{\deg}&\colhead{\deg} &   \colhead{\arcmin} &    \colhead{\arcmin} &          \colhead{K} &       \colhead{\kms} &\colhead{\kms} &     \colhead{ \kms} &         \colhead{$10^{18}$ cm$^{-2}$} &      \colhead{\Msun} }

\startdata
 1 & 18:05:41 &  +11:05:57 &      37.84 &      15.13 &       20.6 &       18.0 &       0.33 &     -177.9 &      -47.6 &       20.8 &            126.8 &       77.4 \\
\hline
 2 & 18:06:28 &  +11:25:00 &      38.22 &      15.09 &        8.8 &        6.0 &       0.29 &     -175.2 &      -43.7 &       22.0 &              20.6 &       12.6 \\
\hline
 3 & 18:06:25 &  +11:38:03 &      38.42 &      15.19 &        8.8 &       12.0 &       0.38 &     -172.5 &      -40.6 &       21.8 &              64.8 &       39.5 \\
\hline
4 & 18:03:51 &  +10:47:21 &      37.35 &      15.40 &       29.5 &       33.0 &       0.37 &     -166.0 &      -37.3 &       25.2 &            324.5 &      197.9 \\
\hline
 5 & 18:01:30 &  +10:26:26 &      36.77 &      15.77 &       29.5 &       24.0 &       0.56 &     -162.6 &      -35.9 &       25.1 &            354.5 &      216.3 \\
\hline
6 &  17:52:23 &  +05:11:35 &      30.84 &      15.49 &       65.7 &       99.0 &       0.57 &     -134.0 &      -25.3 &       30.8 &          2568.4 &     1566.7 \\
\hline
7 &  17:57:37 &  +08:50:35 &      34.84 &      15.94 &       47.4 &       66.0 &       0.75 &     -133.0 &      -12.1 &       28.1 &           1210.0 &      738.1 \\
\hline
 9 & 17:55:32 &  +06:44:08 &      32.64 &      15.48 &       23.8 &       21.0 &       0.39 &     -129.1 &      -14.8 &       28.8 &            257.2 &      156.9 \\
\hline
10 &  17:53:52 &  +13:20:06 &      38.69 &      18.69 &       35.0 &       33.0 &       0.59 &     -130.6 &       -0.3 &       24.6 &            648.4 &      395.5 \\
\hline
11 &  17:34:07 &  +12:08:17 &      35.39 &      22.57 &       11.7 &       12.0 &       0.38 &     -131.0 &      -13.3 &       19.3 &             92.0 &       56.1 \\
\hline
12 &  17:34:24 &  +14:24:58 &      37.68 &      23.44 &       20.3 &       21.0 &       0.65 &     -130.0 &       -6.6 &       21.0 &            253.4 &      154.6 \\
\hline
13 &  17:50:40 &  +03:38:16 &      29.20 &      15.17 &       15.0 &       24.0 &       0.48 &     -129.4 &      -25.8 &       29.9 &           317.3 &      193.6 \\
\hline
14 &  17:51:08 &  +05:14:55 &      30.75 &      15.80 &       12.0 &       12.0 &       0.32 &     -126.6 &      -18.4 &       30.0 &              63.0 &       38.5 \\
\hline
15 & 17:37:02 &  +12:50:36 &      36.40 &      22.22 &       52.7 &       81.0 &       0.76 &     -127.0 &       -6.1 &       26.0 &         2855.4 &     1741.8 \\
\hline
16 &  17:33:05 &  +12:32:49 &      35.68 &      22.97 &       26.4 &       33.0 &       0.42 &     -124.7 &       -6.5 &       22.7 &            346.1 &      211.1 \\
\hline
17 &  17:55:53 &  +07:16:23 &      33.18 &      15.64 &       23.8 &       27.0 &       0.45 &     -123.9 &       -8.0 &       22.2 &            304.1 &      185.5 \\
\hline
 18 & 17:47:49 &  +03:15:59 &      28.52 &      15.63 &       21.0 &       39.0 &       0.35 &     -123.0 &      -21.8 &       22.1 &            263.2 &      160.6 \\
\hline
 19 & 18:02:42 &  +11:19:28 &      37.73 &      15.88 &       26.5 &       24.0 &       0.41 &     -121.1 &        8.4 &       24.0 &            346.1 &      211.1 \\
\hline
20 &  17:51:36 &  +12:45:53 &      37.90 &      18.95 &       43.9 &       63.0 &       0.41 &     -120.7 &        7.1 &       23.6 &           1217.2 &      742.5 \\
\hline
21 &  17:28:21 &  +08:01:20 &      30.71 &      22.10 &       53.5 &       36.0 &       0.45 &     -119.6 &      -15.5 &       24.6 &            637.0 &      388.6 \\
\hline
\multicolumn{ 1}{r}{22} &\multicolumn{ 1}{r}{17:53:45} & \multicolumn{ 1}{r}{+07:30:06} & \multicolumn{ 1}{r}{33.15} & \multicolumn{ 1}{r}{16.22} & \multicolumn{ 1}{r}{23.8} & \multicolumn{ 1}{r}{21.0} & \multicolumn{ 1}{r}{1.38} &     -119.6 &       -4.1 &        5.9 &             85.1 &       51.9 \\
\multicolumn{ 1}{r}{} & \multicolumn{ 1}{r}{} & \multicolumn{ 1}{r}{} & \multicolumn{ 1}{r}{} & \multicolumn{ 1}{r}{} & \multicolumn{ 1}{r}{} & \multicolumn{ 1}{r}{} & \multicolumn{ 1}{r}{} &   -118.3 &       -2.7 &       20.7 &           383.5 &      233.9 \\
\hline
23 &  17:39:37 &  +10:29:11 &      34.39 &      20.66 &      109.2 &      147.0 &       1.61 &     -123.7 &       -7.4 &       25.1 &         9847.3 &     6006.9 \\
\hline
24 &  17:50:52 &  +01:31:45 &      27.29 &      14.16 &       51.0 &       36.0 &       0.64 &     -116.9 &      -19.1 &       31.1 &           731.6 &      446.3 \\
\hline
25 &  17:43:01 &  +13:53:39 &      38.08 &      21.32 &       20.4 &       36.0 &       0.46 &     -117.6 &        8.8 &       21.9 &            462.7 &      282.3 \\
\hline
26 &  17:46:50 &  +03:50:40 &      28.94 &      16.11 &       24.0 &       27.0 &       0.82 &     -117.0 &      -14.7 &       23.6 &            370.0 &      225.7 \\
\hline
27 &  17:37:02 &  +01:34:28 &      25.66 &      17.23 &        9.0 &       15.0 &       0.48 &     -116.4 &      -25.4 &       21.0 &             95.3 &       58.1 \\
\hline
28 &  17:50:28 &  +14:24:37 &      39.37 &      19.88 &       98.8 &       69.0 &       1.03 &     -115.5 &       15.7 &       34.5 &          4563.8 &     2783.9 \\
\hline
29 &  17:34:37 &  +08:52:18 &      32.26 &      21.07 &       14.8 &       21.0 &       0.30 &     -117.3 &       -7.7 &       22.9 &             118.6 &       72.4 \\
\hline
30 &  18:04:43 &  +11:14:24 &      37.87 &      15.40 &       41.2 &       75.0 &       0.80 &     -113.3 &       16.9 &       27.1 &           1352.9 &      825.3 \\
\hline
\multicolumn{ 1}{r}{31} &\multicolumn{ 1}{r}{17:41:33} &  \multicolumn{ 1}{r}{+13:09:57} &      \multicolumn{ 1}{r}{37.21} &     \multicolumn{ 1}{r}{ 21.35} &       \multicolumn{ 1}{r}{11.7} &        \multicolumn{ 1}{r}{9.0} &       \multicolumn{ 1}{r}{0.32} &     -132.0 &       -8.1 &       37.2 &              58.2 &       35.5 \\

\multicolumn{ 1}{r}{} &\multicolumn{ 1}{r}{} & \multicolumn{ 1}{r}{} & \multicolumn{ 1}{r}{} & \multicolumn{ 1}{r}{} & \multicolumn{ 1}{r}{} & \multicolumn{ 1}{r}{} & \multicolumn{ 1}{r}{} &     -111.1 &       12.8 &       14.5 &              26.7 &       16.3 \\
\hline
\multicolumn{ 1}{r}{32} &\multicolumn{ 1}{r}{17:53:15} & \multicolumn{ 1}{r}{+09:54:24} & \multicolumn{ 1}{r}{35.35} & \multicolumn{ 1}{r}{17.38} & \multicolumn{ 1}{r}{345.8} & \multicolumn{ 1}{r}{540.0} & \multicolumn{ 1}{r}{2.65} &     -138.1 &      -16.6 &       24.9 &      17260.0    &    10528.6 \\

\multicolumn{ 1}{r}{} &\multicolumn{ 1}{r}{} & \multicolumn{ 1}{r}{} & \multicolumn{ 1}{r}{} & \multicolumn{ 1}{r}{} & \multicolumn{ 1}{r}{} & \multicolumn{ 1}{r}{} & \multicolumn{ 1}{r}{} &     -110.5 &       11.0 &       30.9 &        59575.7 &    36341.2 \\
\hline
\multicolumn{ 1}{r}{33} &\multicolumn{ 1}{r}{17:50:34} & \multicolumn{ 1}{r}{+13:37:01} & \multicolumn{ 1}{r}{38.61} & \multicolumn{ 1}{r}{19.53} & \multicolumn{ 1}{r}{20.4} & \multicolumn{ 1}{r}{21.0} & \multicolumn{ 1}{r}{0.45} &     -123.9 &        5.4 &       20.0 &             122.0 &       74.4 \\

\multicolumn{ 1}{r}{} &\multicolumn{ 1}{r}{} & \multicolumn{ 1}{r}{} & \multicolumn{ 1}{r}{} & \multicolumn{ 1}{r}{} & \multicolumn{ 1}{r}{} & \multicolumn{ 1}{r}{} & \multicolumn{ 1}{r}{} &  -101.4 &       27.9 &       32.1 &            158.8 &      96.9 \\
\hline
34 &  17:58:05 &  +14:34:51 &      40.33 &      18.27 &       34.8 &       33.0 &       0.53 &     -114.8 &       20.4 &       23.7 &            637.2 &      388.7 \\
\hline
35&  17:34:14 &  +13:48:36 &      37.06 &      23.23 &       26.2 &       27.0 &       0.49 &     -114.2 &        7.7 &       21.1 &            444.3 &      271.0 \\
\hline
36 &  17:52:12 &  +13:37:57 &      38.80 &      19.18 &       35.0 &       48.0 &       0.88 &     -114.9 &       15.3 &       24.9 &            634.5 &      387.1 \\
\hline
\multicolumn{ 1}{r}{ 37} &\multicolumn{ 1}{r}{  17:55:20} &  \multicolumn{ 1}{r}{+05:54:24} &      \multicolumn{ 1}{r}{31.84} &      \multicolumn{ 1}{r}{15.16} &       \multicolumn{ 1}{r}{74.6} &       \multicolumn{ 1}{r}{63.0} &       \multicolumn{ 1}{r}{1.11} &     -121.0 &       -8.9 &       32.7 &          2890.3 &     1763.1 \\

\multicolumn{ 1}{r}{} &\multicolumn{ 1}{r}{} & \multicolumn{ 1}{r}{} & \multicolumn{ 1}{r}{} & \multicolumn{ 1}{r}{} & \multicolumn{ 1}{r}{} & \multicolumn{ 1}{r}{} & \multicolumn{ 1}{r}{} &     -104.8 &        7.2 &       22.2 &          1239.4 &      756.0 \\
\hline
38 &  17:54:45 &  +07:09:09 &      32.94 &      15.84 &        8.9 &       18.0 &       0.30 &     -109.6 &        5.5 &       40.9 &             129.8 &       79.1 \\
\hline
39 &  17:25:35 &  +09:30:07 &      31.83 &      23.36 &       53.3 &       81.0 &       0.62 &     -105.4 &        1.2 &       32.7 &          2372.3 &     1447.1 \\
\hline
40 &  17:49:49 &  +02:33:54 &      28.12 &      14.86 &       45.0 &       51.0 &       0.38 &     -106.8 &       -6.6 &       41.1 &           1572.2 &     959.0 \\
\hline
41 &  18:04:30 &  +14:21:33 &      40.79 &      16.76 &        8.7 &       15.0 &       0.34 &     -106.2 &       31.4 &       20.1 &              47.9 &       29.2 \\
\hline
\multicolumn{ 1}{r}{42} &\multicolumn{ 1}{r}{18:05:42} & \multicolumn{ 1}{r}{+03:00:05} & \multicolumn{ 1}{r}{30.38} & \multicolumn{ 1}{r}{11.54} & \multicolumn{ 1}{r}{15.0} & \multicolumn{ 1}{r}{18.0} & \multicolumn{ 1}{r}{0.58} &     -108.9 &        0.1 &        4.0 &              11.7 &        7.2 \\

\multicolumn{ 1}{r}{} &\multicolumn{ 1}{r}{} & \multicolumn{ 1}{r}{} & \multicolumn{ 1}{r}{} & \multicolumn{ 1}{r}{} & \multicolumn{ 1}{r}{} & \multicolumn{ 1}{r}{} & \multicolumn{ 1}{r}{} &     -103.4 &        5.6 &       26.6 &            92.8 &       56.6 \\
\hline
43 &  17:35:44 &  +02:54:30 &      26.74 &      18.14 &       62.9 &      246.0 &       0.67 &     -104.7 &      -10.6 &       27.4 &         7816.7 &     4768.2 \\
\hline
44 &  17:32:56 &  +01:25:43 &      25.02 &      18.07 &       30.0 &       18.0 &       0.95 &     -104.3 &      -15.9 &       22.7 &            402.2 &      245.3 \\
\hline
45 &  18:03:07 &  +08:22:54 &      35.02 &      14.52 &       14.8 &        9.0 &       0.31 &      -99.1 &       23.1 &       34.4 &            96.0 &       58.5 \\
\hline
46 &  17:29:25 &  +08:35:43 &      31.39 &      22.11 &       50.4 &       24.0 &       0.35 &     -102.7 &        3.5 &       26.3 &            541.4 &      330.2 \\
\hline
47 &  18:04:28 &  +08:04:14 &      34.89 &      14.08 &       14.9 &       21.0 &       0.37 &      -98.1 &       24.0 &       26.0 &             148.6 &      90.7 \\
\hline
48 &  17:57:52 &  +11:48:43 &      37.66 &      17.16 &       14.7 &       21.0 &       0.30 &     -110.9 &       17.6 &       28.5 &             139.1 &       84.9 \\
\hline
49 &  17:41:33 &  +09:40:26 &      33.82 &      19.88 &       23.7 &       30.0 &       0.30 &      -97.1 &       18.1 &       24.3 &            222.7 &      135.8 \\
\hline
50 &  17:26:14 &  +12:28:43 &      34.86 &      24.47 &       14.7 &       18.0 &       0.34 &      -97.0 &       17.5 &       23.9 &             98.3 &       60.0 \\
\hline
51 &  17:46:02 &  +07:13:51 &      32.01 &      17.81 &       23.8 &       24.0 &       0.30 &      -95.9 &       15.2 &       26.2 &            227.6 &      138.8 \\
\hline
52 &  18:09:16 &  +06:37:59 &      34.10 &      12.38 &       17.9 &       24.0 &       0.42 &      -95.9 &       24.6 &       20.9 &            225.8 &      137.8 \\
\hline
53 & 17:43:29 &  +11:33:41 &      35.86 &      20.25 &        8.8 &       24.0 &       0.37 &      -94.6 &       26.3 &       30.2 &            190.3 &      116.1 \\
\hline
54 & 17:51:31 &  +03:45:19 &      29.41 &      15.04 &        9.0 &       12.0 &       0.31 &      -95.3 &        9.0 &       16.0 &              38.2 &       23.3 \\
\hline
55 & 17:45:53 &  +04:12:46 &      29.17 &      16.49 &       23.9 &       45.0 &       0.53 &      -96.4 &        6.4 &       39.3 &           1207.1 &      736.3 \\
\hline
56 &  18:09:39 &  +07:05:16 &      34.56 &      12.50 &       17.9 &       21.0 &       0.38 &      -95.7 &       26.2 &       24.2 &             144.4 &      88.1 \\
\hline
57 &  17:42:13 &  +03:55:11 &      28.46 &      17.17 &       38.9 &       39.0 &       0.56 &      -93.4 &        6.7 &       25.3 &           980.6 &      587.2 \\
\hline
58 &  17:27:43 &  +09:25:07 &      32.00 &      22.84 &        8.9 &       15.0 &       0.30 &      -92.5 &       14.9 &       19.7 &              34.4 &       21.0 \\
\hline
59 & 18:10:42 &  +06:41:31 &      34.32 &      12.09 &       17.9 &       18.0 &       0.31 &      -90.0 &       31.2 &       25.6 &             158.6 &      96.7 \\
\hline
60 &  18:08:20 &  +11:30:05 &      38.50 &      14.71 &       11.8 &       15.0 &       0.50 &      -90.4 &       42.1 &       17.4 &             106.8 &       65.2 \\
\hline
62 &  18:06:04 &  +04:25:39 &      31.72 &      12.10 &        9.0 &        9.0 &       0.30 &      -87.7 &       25.4 &       17.2 &             24.0 &       14.7 \\
\hline
65 &  18:05:16 &  +07:29:11 &      34.44 &      13.65 &       17.9 &       18.0 &       0.69 &      -85.7 &       35.2 &       17.7 &            225.8 &      137.7 \\
\hline
66 &  17:58:15 &  +06:45:39 &      32.97 &      14.89 &        6.0 &       12.0 &       0.31 &      -86.3 &       29.4 &       16.5 &              28.0 &       17.1 \\
\hline
\multicolumn{ 1}{r}{67} &\multicolumn{ 1}{r}{  17:59:10} &  \multicolumn{ 1}{r}{+02:50:57} &      \multicolumn{ 1}{r}{29.48} &      \multicolumn{ 1}{r}{12.92} &       \multicolumn{ 1}{r}{15.0} &       \multicolumn{ 1}{r}{12.0} &       \multicolumn{ 1}{r}{0.47} &      -86.5 &       19.0 &       14.5 &         91.5 &       55.8 \\

\multicolumn{ 1}{r}{} &\multicolumn{ 1}{r}{} & \multicolumn{ 1}{r}{} & \multicolumn{ 1}{r}{} & \multicolumn{ 1}{r}{} & \multicolumn{ 1}{r}{} & \multicolumn{ 1}{r}{} & \multicolumn{ 1}{r}{} &      -74.5 &       31.0 &       69.6 &     96.0 &       58.5 \\
\hline
68 &  17:49:49 &  +04:33:20 &      29.95 &      15.78 &       12.0 &       48.0 &       0.33 &      -84.8 &       20.9 &       22.9 &            241.2 &      147.1 \\
\hline
69 &  17:29:29 &  +09:47:39 &      32.57 &      22.62 &        5.9 &        6.0 &       0.31 &      -91.0 &       18.3 &       24.5 &              25.7 &       15.7 \\
\hline
70 &  17:41:24 &  +10:53:06 &      34.97 &      20.43 &       23.6 &       21.0 &       0.31 &      -82.8 &       35.3 &       24.6 &           177.2 &      108.1 \\
\hline
71 &  18:06:09 &  +11:39:44 &      38.42 &      15.26 &       11.8 &        9.0 &       0.30 &      -84.4 &       47.5 &       28.8 &             101.1 &       61.7 \\
\hline
72 &  18:00:12 &  +03:28:21 &      30.17 &      12.97 &       18.0 &       21.0 &       0.30 &      -80.1 &       27.6 &       25.1 &            181.0 &      110.4 \\
\hline
73 & 18:01:02 &  +04:22:56 &      31.10 &      13.20 &        9.0 &       12.0 &       0.34 &      -81.4 &       29.2 &       18.1 &              55.4 &       33.8 \\
\hline
74 &  18:01:39 &  +07:58:18 &      34.48 &      14.66 &       11.9 &       27.0 &       0.43 &      -79.2 &       41.3 &       22.7 &            191.4 &      116.7 \\
\hline
75 &  18:03:44 &  +07:50:30 &      34.59 &      14.15 &       14.9 &       12.0 &       0.31 &      -78.5 &       42.6 &       23.7 &             110.2 &       67.2 \\
\hline
77 &  18:02:07 &  +03:36:21 &      30.51 &      12.61 &       15.0 &       15.0 &       0.34 &      -77.2 &       31.8 &       27.4 &             128.5 &       78.4 \\
\hline
80 &  17:36:30 &  +01:31:18 &      25.54 &      17.33 &       30.0 &       27.0 &       0.38 &      -74.1 &       16.5 &       27.7 &            253.0 &      154.3 \\
\hline
81 &  18:04:27 &  +07:15:38 &      34.14 &      13.73 &       17.9 &       18.0 &       0.53 &      -76.3 &       43.6 &       20.4 &            176.1 &      107.4 \\
\hline
84 &  18:00:57 &  +05:13:15 &      31.86 &      13.60 &       56.8 &      111.0 &       0.70 &      -70.9 &       41.9 &       26.6 &           1643.8 &     1002.7 \\
\hline
87 &  17:46:25 &  +04:38:56 &      29.64 &      16.57 &       15.0 &       27.0 &       0.39 &      -69.1 &       35.2 &       29.0 &            242.4 &      147.9 \\
\hline
\multicolumn{ 1}{r}{88} &\multicolumn{ 1}{r}{17:56:59} & \multicolumn{ 1}{r}{+10:12:34} & \multicolumn{ 1}{r}{36.05} & \multicolumn{ 1}{r}{16.68} & \multicolumn{ 1}{r}{44.3} & \multicolumn{ 1}{r}{45.0} & \multicolumn{ 1}{r}{2.23}&     -70.5 &       53.5 &       33.2 &           985.6 &      601.2 \\

\multicolumn{ 1}{r}{} &\multicolumn{ 1}{r}{} & \multicolumn{ 1}{r}{} & \multicolumn{ 1}{r}{} & \multicolumn{ 1}{r}{} & \multicolumn{ 1}{r}{} & \multicolumn{ 1}{r}{} & \multicolumn{ 1}{r}{} &   -66.6 &       57.4 &       11.1 &            756.7 &      461.6 \\
\hline
89 &  17:34:13 &  +11:03:10 &      34.34 &      22.10 &       64.8 &      111.0 &       0.51 &      -65.6 &       49.4 &       40.0 &          3603.7 &     2198.3 \\
\hline
90 &  18:03:41 &  +13:50:53 &      40.22 &      16.73 &       14.6 &       12.0 &       0.33 &      -68.2 &       67.9 &       21.5 &              49.1 &       29.9 \\
  \enddata
 \tablenotetext{a} {\scriptsize At the distance of 10kpc, 1\arcmin = 2.9 pc.}
 \tablenotetext{b} {\scriptsize At the distance of 10 kpc.}
 \label{Ctab}
\end{deluxetable}

\clearpage
\begin{deluxetable}{rrrrrrrrrrrrr}
\rotate
\tabletypesize{\scriptsize}
\tablewidth{0pc}
\tablecaption{Catalog of Clouds in the Complex C Data with $-65 <$  \vlsr $< -50$\kms. }

\tablehead{  \colhead{Cloud \#} &\colhead{RA} &  \colhead{DEC} & \colhead{l} & \colhead{b} &\colhead{$\Delta RA$\tablenotemark{a}} & \colhead{$\Delta Dec$\tablenotemark{a}} & \colhead{$T_{b,peak}$} &  \colhead{\vlsr} & \colhead{\vgsr}& \colhead{$\Delta V$} &  \colhead{\NH(tot)} & \colhead{\Mhi\tablenotemark{b}} \\
&  \colhead{(J2000)} &    \colhead{(J2000)} & \colhead{\deg}& \colhead{\deg}&   \colhead{\arcmin} &    \colhead{\arcmin} &          \colhead{K} &       \colhead{\kms} &\colhead{\kms} &     \colhead{ \kms} &         \colhead{$10^{18}$ cm$^{-2}$} &      \colhead{\Msun} }

\startdata

91 &  18:08:03 &  +03:59:08 &      31.55 &      11.46 &       21.0 &       21.0 &       0.45 &      -63.7 &       49.1 &       29.8 &          274.4 &     167.4 \\
\hline
94 &  17:53:28 &  +14:31:47 &      39.80 &      19.27 &       43.6 &       48.0 &       0.55 &      -62.3 &       70.6 &       29.6 &          1769.6 &    1079.5 \\
\hline
98 &  17:58:41 &  +11:18:22 &      37.27 &      16.77 &       11.8 &       24.0 &       0.36 &      -59.9 &       67.6 &       26.7 &          171.4 &     104.5 \\
\hline
101 &  17:46:21 &  +08:34:07 &      33.31 &      18.33 &       14.8 &       21.0 &       0.30 &      -59.8 &       54.9 &       24.4 &            109.8 &      67.0 \\
\hline
\multicolumn{ 1}{r}{  104 } &\multicolumn{ 1}{r}{  17:58:52 } &  \multicolumn{ 1}{r}{+14:12:25 } &  \multicolumn{ 1}{r}{ 40.05 } &  \multicolumn{ 1}{r}{17.94 } &  \multicolumn{ 1}{r}{ 26.2 } &  \multicolumn{ 1}{r}{   36.0 } &  \multicolumn{ 1}{r}{    0.42} &      -61.4 &       73.3 &       19.5 &         405.4 &     247.3 \\

\multicolumn{ 1}{r}{} &\multicolumn{ 1}{r}{} & \multicolumn{ 1}{r}{} & \multicolumn{ 1}{r}{} & \multicolumn{ 1}{r}{} & \multicolumn{ 1}{r}{} & \multicolumn{ 1}{r}{} & \multicolumn{ 1}{r}{} &     -53.0 &       81.7 &        4.7 &           21.0 &      12.8 \\
\hline
106 &  18:04:33 &  +06:30:35 &      33.45 &      13.37 &       47.7 &       60.0 &       0.78 &      -59.6 &       58.4 &       24.9 &          1113.8 &     679.4 \\
\hline
108 &  17:41:04 &  +06:42:11 &      30.94 &      18.69 &       56.6 &       51.0 &       0.68 &      -58.9 &       48.2 &       31.1 &        1711.4 &    1044.0 \\
\hline
109 &  18:02:30 &  +14:19:41 &      40.55 &      17.19 &       26.2 &       39.0 &       0.36 &      -59.4 &       77.3 &       27.8 &           603.3 &     368.0 \\
\hline
110 &  18:07:13 &  +12:55:02 &      39.71 &      15.56 &       11.7 &       12.0 &       0.44 &      -59.8 &       75.6 &       22.9 &          99.6 &      60.8 \\
\hline
 \multicolumn{ 1}{r}{111} & \multicolumn{ 1}{r}{ 17:47:40 } &  \multicolumn{ 1}{r}{ +03:46:08} &  \multicolumn{ 1}{r}{  28.97} &  \multicolumn{ 1}{r}{  15.89 } &  \multicolumn{ 1}{r}{26.9 } &  \multicolumn{ 1}{r}{    27.0} &  \multicolumn{ 1}{r}{   0.57} &      -61.1 &       41.4 &       19.0 &             164.0 &     100.1 \\

\multicolumn{ 1}{r}{} &\multicolumn{ 1}{r}{} & \multicolumn{ 1}{r}{} & \multicolumn{ 1}{r}{} & \multicolumn{ 1}{r}{} & \multicolumn{ 1}{r}{} & \multicolumn{ 1}{r}{} & \multicolumn{ 1}{r}{} &      -54.7 &       47.8 &        7.1 &         45.5 &      27.7 \\
\hline
112 &  18:10:17 &  +11:04:13 &      38.32 &      14.09 &       35.3 &       84.0 &       2.85 &      -60.7 &       71.6 &       12.0 &           869.3 &     530.3 \\
\hline
114 & 17:49:47 &  +14:38:05 &      39.51 &      20.12 &       34.8 &       21.0 &       0.41 &      -57.6 &       73.9 &       20.3 &          335.8 &     204.8 \\
\hline
117 & 17:20:45 &  +12:16:24 &      34.04 &      25.60 &        8.8 &        9.0 &       0.35 &      -57.6 &       53.5 &       24.9 &             58.6 &      35.8 \\
  \enddata
 \tablenotetext{a} {\scriptsize At the distance of 10kpc, 1\arcmin = 2.9 pc.}
 \tablenotetext{b} {\scriptsize At the distance of 10 kpc.}
 \label{Ctab2}
\end{deluxetable}

\clearpage
\begin{deluxetable}{rrrrrrrrrrrrr}
\rotate
\tabletypesize{\scriptsize}
\tablewidth{0pc}
\tablecaption{Smoothed Complex C Cloud Catalog}
\tablehead{\colhead{Cloud \#} &  \colhead{RA} &  \colhead{DEC} & \colhead{l} & \colhead{b} &\colhead{$\Delta RA$\tablenotemark{a}} & \colhead{$\Delta Dec$\tablenotemark{a}} & \colhead{$T_{b,peak}$} &  \colhead{\vlsr} & \colhead{\vgsr}& \colhead{$\Delta V$} &  \colhead{\NH(tot)} & \colhead{\Mhi\tablenotemark{b}} \\
 & \colhead{(J2000)} &    \colhead{(J2000)} & \colhead{\deg}& \colhead{\deg}&   \colhead{\arcmin} &    \colhead{\arcmin} &          \colhead{K} &       \colhead{\kms} &\colhead{\kms} &     \colhead{ \kms} &         \colhead{$10^{18}$ cm$^{-2}$} &      \colhead{\Msun} }

\startdata
1 &17:57:33 & +08:50:39 &      34.83 &      15.96 &       41.5 &       63.0 &       0.16 &     -130.4 &       -9.6 &       36.4 &           1523.4 &     929.2 \\
\hline
2 & 17:33:05 & +12:29:59 &      35.63 &      22.95 &       26.4 &       42.0 &       0.14 &     -126.2 &       -8.2 &       22.3 &            312.0 &      190.3 \\
\hline
3 & 17:51:04 & +01:28:43 &      27.27 &      14.09 &       33.0 &       24.0 &       0.18 &     -120.4 &      -22.6 &       26.2 &            328.6 &      200.5 \\
\hline
4 &17:28:24 & +08:02:01 &      30.73 &      22.09 &       29.7 &       33.0 &       0.15 &     -119.9 &      -15.7 &       26.0 &            478.3 &      291.7 \\
\hline
\multicolumn{ 1}{r}{5}  &\multicolumn{ 1}{r}{17:54:13} & \multicolumn{ 1}{r}{+07:26:50} & \multicolumn{ 1}{r}{33.15} & \multicolumn{ 1}{r}{16.09} & \multicolumn{ 1}{r}{59.5} & \multicolumn{ 1}{r}{45.0} & \multicolumn{ 1}{r}{0.32} &     -119.2 &       -3.6 &       26.0 &            840.0 &      512.4 \\

\multicolumn{ 1}{r}{} &\multicolumn{ 1}{r}{} & \multicolumn{ 1}{r}{} & \multicolumn{ 1}{r}{} & \multicolumn{ 1}{r}{} & \multicolumn{ 1}{r}{} & \multicolumn{ 1}{r}{} & \multicolumn{ 1}{r}{} &     -119.5 &       -4.0 &        5.9 &     92.0 &       56.1 \\
\hline
6 &17:42:54 & +13:53:17 &      38.06 &      21.34 &       29.1 &       45.0 &       0.17 &     -116.3 &       10.0 &       23.4 &            581.6 &      354.8 \\
\hline
7 &17:34:13 & +13:50:47 &      37.09 &      23.25 &       26.2 &       36.0 &       0.17 &     -114.2 &        7.7 &       21.9 &            437.5 &      266.8 \\
\hline
8 & 17:49:53 & +02:35:52 &      28.15 &      14.87 &       53.9 &       72.0 &       0.17 &     -106.7 &       -6.3 &       44.1 &          1997.3 &     1218.4 \\
\hline
9 &17:25:38 & +09:33:24 &      31.89 &      23.37 &       56.2 &       78.0 &       0.23 &     -105.0 &        1.7 &       33.5 &         2485.2 &     1516.0 \\
\hline
10 & 17:35:24 & +02:52:32 &      26.67 &      18.20 &      116.9 &      249.0 &       0.37 &     -105.1 &      -11.3 &       28.5 &        9516.7 &     5805.2 \\
\hline
11 & 17:38:14 & +10:48:51 &      34.55 &      21.10 &      170.9 &      255.0 &       0.81 &     -124.1 &       -7.7 &       26.9 &        15145.5 &    9238.8 \\
\hline
12 &17:55:11 & +09:34:30 &      35.25 &      16.80 &      434.9 &      633.0 &       1.08 &     -113.0 &        8.6 &       45.0 &       117964.9 &    71958.6 \\
\hline
13 & 17:42:23 & +03:54:32 &      28.47 &      17.13 &       38.9 &       45.0 &       0.21 &      -94.6 &        5.7 &       24.9 &         903.7 &      551.2 \\
\hline
14 & 17:46:08 & +02:39:50 &      27.77 &      15.73 &       42.0 &       45.0 &       0.14 &      -95.4 &        3.3 &       27.0 &          671.3 &      4009.5 \\
\hline
15 & 18:04:59 & +07:28:14 &      34.39 &      13.70 &       29.8 &       36.0 &       0.14 &      -84.2 &       36.6 &       19.1 &         244.7 &      149.3 \\
\hline
\multicolumn{ 1}{r}{16} &\multicolumn{ 1}{r}{18:00:37} & \multicolumn{ 1}{r}{+04:55:21} & \multicolumn{ 1}{r}{31.54} & \multicolumn{ 1}{r}{13.54} & \multicolumn{ 1}{r}{53.8} & \multicolumn{ 1}{r}{111.0} & \multicolumn{ 1}{r}{0.20} &      -80.7 &       31.2 &       21.3 &         1007.3 &      614.4 \\

\multicolumn{ 1}{r}{} &\multicolumn{ 1}{r}{} & \multicolumn{ 1}{r}{} & \multicolumn{ 1}{r}{} & \multicolumn{ 1}{r}{} & \multicolumn{ 1}{r}{} & \multicolumn{ 1}{r}{} & \multicolumn{ 1}{r}{} &      -67.7 &        9.0 &       13.3 &           381.6 &      232.8 \\
 \enddata
 \tablenotetext{a} {\scriptsize At the distance of 10kpc, 1\arcmin = 2.9 pc.}
 \tablenotetext{b} {\scriptsize At the distance of 10 kpc.}
 \label{CStab}
\end{deluxetable}

\clearpage
\begin{deluxetable}{rrrrrrrrrrrrr}
\rotate
\tabletypesize{\scriptsize}
\tablewidth{0pc}
\tablecaption{Magellanic Stream Cloud Catalog}
\tablehead{\colhead{Cloud \#} &  \colhead{RA} &  \colhead{DEC} & \colhead{l} & \colhead{b} &\colhead{$\Delta RA$\tablenotemark{a}} & \colhead{$\Delta Dec$\tablenotemark{a}} & \colhead{$T_{b,peak}$} &  \colhead{\vlsr} & \colhead{\vgsr}& \colhead{$\Delta V$} &  \colhead{\NH(tot)} & \colhead{\Mhi\tablenotemark{b}} \\
&  \colhead{(J2000)} &    \colhead{(J2000)} &\colhead{\deg} &\colhead{\deg} &   \colhead{\arcmin} &    \colhead{\arcmin} &          \colhead{K} &       \colhead{\kms} &\colhead{\kms} &     \colhead{ \kms} &         \colhead{$10^{18}$ cm$^{-2}$} &      \colhead{\Msun} }

\startdata

2 &  22:23:33 &  +20:26:16 &      82.10 &     -30.47 &       14.1 &       21.0 &       0.39 &     -411.0 &     -223.1 &       18.5 &          84.4 &     1852.9 \\
\hline
5 &  22:08:53 &  +17:15:38 &      76.64 &     -30.64 &       25.8 &       24.0 &       0.34 &     -407.8 &     -223.7 &       25.5 &       200.0 &     4391.0 \\
\hline
7 &  22:19:13 &  +21:04:50 &      81.65 &     -29.33 &       33.6 &       27.0 &       0.34 &     -406.5 &     -216.7 &       20.7 &       223.8 &     4914.4 \\
\hline
8 &  22:42:31 &  +18:41:13 &      85.10 &     -34.55 &       25.6 &       18.0 &       0.28 &     -407.2 &     -226.7 &       26.2 &         180.4 &     3962.0 \\
\hline
9 &  23:25:38 &  +18:41:18 &      96.18 &     -39.69 &       17.1 &       24.0 &       0.28 &     -404.5 &     -236.2 &       19.5 &           119.0 &     2612.9 \\
\hline
11 &  22:24:20 &  +20:05:59 &      82.02 &     -30.85 &       19.7 &       30.0 &       0.52 &     -403.7 &     -216.6 &       23.6 &         384.9 &     8451.6 \\
\hline
13 &  22:17:21 &  +21:10:09 &      81.33 &     -28.99 &       11.2 &       15.0 &       0.30 &     -400.9 &     -210.7 &       25.1 &          57.9 &     1270.8 \\
\hline
 \multicolumn{ 1}{r}{14} & \multicolumn{ 1}{r}{ 22:40:05} & \multicolumn{ 1}{r}{ +17:10:27} & \multicolumn{ 1}{r}{83.45} & \multicolumn{ 1}{r}{-35.44} & \multicolumn{ 1}{r}{  11.5} & \multicolumn{ 1}{r}{ 15.0} & \multicolumn{ 1}{r}{0.35} &     -399.7 &     -221.6 &       61.1 &       120.0 &     2636.5 \\

\multicolumn{ 1}{r}{} &\multicolumn{ 1}{r}{} & \multicolumn{ 1}{r}{} & \multicolumn{ 1}{r}{} & \multicolumn{ 1}{r}{} & \multicolumn{ 1}{r}{} & \multicolumn{ 1}{r}{} & \multicolumn{ 1}{r}{} &     -401.8 &     -223.7 &        9.2 &        20.8 &      456.8 \\
\hline
15 &  23:04:18 &  +20:35:35 &      91.64 &     -35.69 &       42.1 &       57.0 &       0.84 &     -402.5 &     -223.9 &       38.6 &      1971.7 &    43297.7 \\
\hline
17 &  22:30:03 &  +19:22:30 &      82.75 &     -32.25 &        8.5 &        9.0 &       0.28 &     -401.2 &     -216.6 &       28.0 &             36.4 &      799.2 \\
\hline
18 &  23:00:03 &  +17:17:34 &      88.46 &     -38.01 &       14.3 &       18.0 &       0.30 &     -396.8 &     -223.5 &       25.9 &          89.0 &     1954.8 \\
\hline
19 &  22:26:05 &  +17:37:57 &      80.59 &     -33.03 &       22.9 &       33.0 &       0.28 &     -401.3 &     -219.3 &       30.0 &          209.0 &     4589.1 \\
\hline
20 &  22:29:32 &  +17:04:13 &      80.93 &     -33.99 &       22.9 &       18.0 &       0.37 &     -395.7 &     -215.6 &       29.5 &        261.6 &     5744.5 \\
\hline
21 &  22:44:44 &  +18:44:06 &      85.66 &     -34.81 &       22.7 &       33.0 &       0.33 &     -391.6 &     -211.5 &       44.8 &         458.0 &    10058.1 \\
\hline
22 &  22:45:27 &  +21:08:44 &      87.46 &     -32.91 &       14.0 &       24.0 &       0.36 &     -389.6 &     -205.1 &       30.5 &         96.7 &     2122.6 \\
\hline
28 &  22:44:39 &  +21:18:30 &      87.38 &     -32.67 &       11.2 &       12.0 &       0.30 &     -385.9 &     -200.9 &       25.6 &           36.6 &      804.0 \\
\hline
29 &  22:14:14 &  +21:44:48 &      81.11 &     -28.08 &        5.6 &        6.0 &       0.31 &     -384.9 &     -193.1 &        9.3 &            9.8 &      215.4 \\
\hline
31 &  22:14:41 &  +20:50:40 &      80.55 &     -28.84 &       14.0 &       15.0 &       0.29 &     -385.1 &     -195.0 &       31.5 &         102.9 &     2260.4 \\
\hline
33 &  23:10:07 &  +19:13:22 &      92.30 &     -37.55 &       39.7 &       27.0 &       0.78 &     -381.6 &     -207.3 &       20.4 &        448.1 &    9839.6 \\
\hline
34 &  22:30:32 &  +18:24:46 &      82.16 &     -33.08 &       17.1 &       12.0 &       0.32 &     -382.1 &     -199.5 &       20.4 &         63.4 &     1391.6 \\
\hline
38 &  23:13:56 &  +16:20:22 &      91.54 &     -40.52 &       11.5 &       12.0 &       0.27 &     -376.0 &     -208.8 &       24.9 &           64.0 &     1404.7 \\
\hline
39 &  22:57:09 &  +16:55:47 &      87.47 &     -37.95 &       23.0 &       15.0 &       0.39 &     -376.4 &     -203.1 &       26.4 &        195.2 &     4286.5 \\
\hline
43 &  22:37:10 &  +17:47:25 &      83.21 &     -34.53 &       20.0 &       18.0 &       0.29 &     -372.8 &     -192.8 &       22.4 &         65.1 &     1428.5 \\
\hline
44 &  22:44:35 &  +19:10:48 &      85.93 &     -34.42 &       14.2 &       15.0 &       0.29 &     -378.1 &     -197.1 &       23.1 &          58.2 &     1277.2 \\
\hline
45 &  22:29:42 &  +16:37:00 &      80.63 &     -34.37 &       25.9 &       21.0 &       0.31 &     -371.0 &     -191.9 &       27.7 &        230.8 &     5067.5 \\
\hline
47 &  22:51:11 &  +18:38:14 &      87.14 &     -35.74 &       14.2 &       18.0 &       0.52 &     -369.5 &     -191.1 &       27.8 &            157.7 &     3463.4 \\
\hline
48 &  22:47:14 &  +18:57:50 &      86.41 &     -34.95 &        8.5 &       12.0 &       0.34 &     -367.2 &     -187.2 &       17.6 &        28.6 &      627.1 \\
\hline
51 &  23:14:49 &  +17:29:52 &      92.50 &     -39.60 &       77.3 &       54.0 &       0.48 &     -362.6 &     -193.2 &       34.9 &       2131.7 &    46811.1 \\
\hline
52 &  23:03:57 &  +16:35:43 &      89.01 &     -39.09 &       31.6 &       27.0 &       0.63 &     -358.8 &     -188.1 &       30.4 &        559.8 &    12292.4 \\
\hline
54 &  22:41:26 &  +16:25:22 &      83.22 &     -36.23 &       25.9 &       15.0 &       0.44 &     -357.9 &     -181.7 &       25.1 &          215.5 &     4731.6 \\
\hline
57 &  23:25:00 &  +17:00:53 &      95.06 &     -41.13 &       14.3 &       18.0 &       0.43 &     -353.6 &     -188.6 &       22.2 &           143.2 &     3144.2 \\
\hline
59 &  23:15:25 &  +16:31:48 &      92.07 &     -40.52 &       28.8 &       27.0 &       0.48 &     -345.8 &     -178.7 &       38.3 &          458.7 &    10073.3 \\
\hline
67 &  23:14:11 &  +16:52:08 &      91.94 &     -40.08 &       20.1 &       21.0 &       0.31 &     -333.2 &     -165.0 &       28.9 &          188.0 &     4129.3 \\
\hline
76 &  21:59:51 &  +20:46:43 &      77.55 &     -26.60 &       36.5 &       54.0 &       0.48 &     -307.1 &     -115.0 &       27.0 &      1015.4 &    22298.1 \\
\hline
2 &  23:49:08 &  +11:45:24 &      99.67 &     -48.28 &       17.6 &       12.0 &       0.25 &     -391.0 &     -246.7 &       37.1 &        157.1 &     3449.2 \\
\hline
3 &  22:49:24 &  +14:15:31 &      83.58 &     -39.11 &       81.4 &      105.0 &       0.41 &     -387.5 &     -217.8 &       41.6 &      3849.9 &    84543.9 \\
\hline
\multicolumn{ 1}{r}{4} &\multicolumn{ 1}{r}{  22:46:28} & \multicolumn{ 1}{r}{  +15:04:24 } & \multicolumn{ 1}{r}{    83.45} & \multicolumn{ 1}{r}{    -38.04} & \multicolumn{ 1}{r}{       8.7 } & \multicolumn{ 1}{r}{      15.0 } & \multicolumn{ 1}{r}{     0.18} &     -375.4 &     -203.3 &       39.2 &           62.7 &     1376.7 \\
\multicolumn{ 1}{r}{} &\multicolumn{ 1}{r}{} & \multicolumn{ 1}{r}{} & \multicolumn{ 1}{r}{} & \multicolumn{ 1}{r}{} & \multicolumn{ 1}{r}{} & \multicolumn{ 1}{r}{} & \multicolumn{ 1}{r}{} &  -351.6 &     -179.5 &       21.3 &          20.3 &      445.7 \\
\hline
\multicolumn{ 1}{r}{6} &\multicolumn{ 1}{r}{ 22:58:46 }& \multicolumn{ 1}{r}{ +14:51:49} &      \multicolumn{ 1}{r}{86.44} &  \multicolumn{ 1}{r}{   -39.89 }&   \multicolumn{ 1}{r}{    29.0}&       \multicolumn{ 1}{r}{18.0 }&     \multicolumn{ 1}{r}{  0.16 }&     -367.9 &     -199.5 &       50.9 &            182.1 &     3999.9 \\
\multicolumn{ 1}{r}{} &\multicolumn{ 1}{r}{} & \multicolumn{ 1}{r}{} & \multicolumn{ 1}{r}{} & \multicolumn{ 1}{r}{} & \multicolumn{ 1}{r}{} & \multicolumn{ 1}{r}{} & \multicolumn{ 1}{r}{} &
   -376.7 &     -208.3 &       15.7 &       23.8 &      523.4 \\
\hline
7 & 23:06:39 &  +12:43:00 &      87.04 &     -42.74 &      125.8 &      225.0 &       0.60 &     -375.1 &     -213.7 &       45.0 &        12175.2 &   267368.4 \\
\hline
8 &  22:52:57 &  +12:22:29 &      83.06 &     -41.15 &       61.5 &       93.0 &       0.38 &     -373.3 &     -208.8 &       33.0 &     2684.2 &    58944.7 \\
\hline
9 &  22:53:25 &  +13:13:17 &      83.83 &     -40.52 &       20.4 &       24.0 &       0.23 &     -372.9 &     -206.6 &       30.7 &        140.3 &     3081.5 \\
\hline
10 &  23:06:41 &  +15:20:24 &      88.89 &     -40.51 &       37.6 &       27.0 &       0.21 &     -372.2 &     -205.0 &       34.0 &          421.9 &    9264.1 \\
\hline
11 &  23:02:06 &  +15:37:08 &      87.85 &     -39.69 &       14.5 &       12.0 &       0.16 &     -366.8 &     -197.6 &       29.4 &          53.6 &     1177.6 \\
\hline
12 &  22:31:07 &  +15:54:39 &      80.42 &     -35.13 &       20.2 &       15.0 &       0.20 &     -364.6 &     -187.1 &       27.6 &       100.1 &     2198.7 \\
\hline
13 &  23:03:15 &  +13:14:02 &      86.47 &     -41.85 &       58.4 &       72.0 &       0.26 &     -362.2 &     -198.6 &       31.6 &         879.1 &    19304.8 \\
\hline
14 &  23:45:14 &  +14:53:51 &     100.02 &     -45.01 &       11.6 &       27.0 &       0.16 &     -357.5 &     -204.3 &       31.3 &        68.4 &     1501.0 \\
\hline
15 &  22:46:25 &  +15:25:13 &      83.70 &     -37.75 &       31.8 &       51.0 &       0.24 &     -359.1 &     -186.2 &       31.7 &         519.2 &    11401.1 \\
\hline
16 &  22:55:41 &  +15:09:05 &      85.84 &     -39.24 &       20.3 &       21.0 &       0.16 &     -357.2 &     -187.2 &       25.6 &       110.6 &     2429.7 \\
\hline
17 &  23:35:14 &  +14:15:21 &      96.52 &     -44.67 &       20.4 &       15.0 &       0.21 &     -356.6 &     -201.2 &       24.8 &        126.1 &     2768.8 \\
\hline
18 &  22:37:07 &  +14:08:33 &      80.46 &     -37.42 &       20.4 &       21.0 &       0.38 &     -355.3 &     -182.9 &       24.2 &         183.0 &     4018.8 \\
\hline
19 &  23:34:07 &  +13:06:13 &      95.49 &     -45.61 &       35.1 &       21.0 &       0.32 &     -353.9 &     -200.7 &       28.3 &          269.8 &     5924.8 \\
\hline
21 &  23:00:10 &  +14:20:36 &      86.44 &     -40.51 &       23.3 &       33.0 &       0.36 &     -355.5 &     -188.6 &       29.0 &        380.7 &     8359.3 \\
\hline
22 &  22:25:12 &  +14:42:57 &      78.14 &     -35.14 &       20.3 &       24.0 &       0.19 &     -353.2 &     -177.1 &       23.6 &       185.4 &     4070.6 \\
\hline
23 &  23:21:19 &  +13:04:41 &      91.54 &     -44.23 &       14.6 &       15.0 &       0.16 &     -354.4 &     -196.8 &       27.3 &         62.2 &     1366.5 \\
\hline
24 &  23:44:47 &  +12:19:33 &      98.50 &     -47.36 &       11.7 &        9.0 &       0.16 &     -353.8 &     -206.4 &       22.3 &          31.8 &     699.1 \\
\hline
25 &  23:45:38 &  +15:21:27 &     100.38 &     -44.61 &       20.3 &       39.0 &       0.29 &     -351.9 &     -197.9 &       24.8 &         240.3 &     5276.8 \\
\hline
26 &  23:12:55 &  +14:04:30 &      89.77 &     -42.36 &      122.2 &      111.0 &       0.48 &     -352.0 &     -189.5 &       44.5 &      3624.9 &    79603.2 \\
\hline
27 &  23:33:27 &  +12:42:45 &      95.05 &     -45.89 &       41.0 &       24.0 &       0.19 &     -349.8 &     -197.2 &       32.1 &        248.5 &     5456.9 \\
\hline
29 &  22:55:56 &  +13:41:37 &      84.85 &     -40.49 &       14.6 &       15.0 &       0.26 &     -349.8 &     -183.1 &       22.9 &          83.1 &     1824.3 \\
\hline
30 &  23:01:40 &  +12:35:50 &      85.57 &     -42.18 &       41.0 &       27.0 &       0.35 &     -352.3 &     -189.8 &       41.1 &        464.0 &    10188.5 \\
\hline
31 &  23:36:31 &  +12:45:56 &      96.05 &     -46.16 &       11.7 &       12.0 &       0.21 &     -351.2 &     -199.6 &       25.8 &        41.0 &     900.8 \\
\hline
33 &  23:02:18 &  +11:48:05 &      85.14 &     -42.92 &       35.2 &       18.0 &       0.38 &     -349.0 &     -188.5 &       24.9 &       364.3 &     8000.0 \\
\hline
34 &  23:11:40 &  +12:58:50 &      88.65 &     -43.15 &       23.4 &       24.0 &       0.42 &     -346.6 &     -186.2 &       20.2 &        184.1 &     4043.6 \\
\hline
35 &  23:25:26 &  +14:05:38 &      93.43 &     -43.80 &       87.3 &       78.0 &       0.29 &     -346.3 &     -187.8 &       31.8 &       1487.7 &    32699.9 \\
\hline
37 &  22:58:00 &  +12:22:12 &      84.40 &     -41.86 &       29.3 &       24.0 &       0.17 &     -341.4 &     -178.3 &       27.0 &         158.1 &     3471.8 \\
\hline
38 &  23:26:30 &  +12:12:48 &      92.55 &     -45.59 &      202.3 &      102.0 &       0.65 &     -341.4 &     -187.6 &       30.3 &        6359.7 &   139659.1 \\
\hline
43 &  22:38:41 &  +13:54:50 &      80.66 &     -37.83 &       20.4 &       27.0 &       0.22 &     -339.2 &     -167.7 &       34.9 &          253.9 &     5576.4 \\
\hline
49 &  22:33:25 &  +13:17:41 &      78.91 &     -37.51 &       55.5 &       39.0 &       0.26 &     -328.2 &     -156.9 &       27.7 &        964.1 &    21172.5 \\
\hline
50 &  22:38:01 &  +14:31:34 &      80.97 &     -37.25 &       29.0 &       21.0 &       0.20 &     -328.3 &     -155.3 &       28.0 &        221.7 &     4868.3 \\
\hline
51 &  23:17:54 &  +14:50:37 &      91.70 &     -42.28 &       17.4 &       27.0 &       0.18 &     -328.1 &     -165.4 &       23.0 &           81.3 &     1785.8 \\
\hline
\multicolumn{ 1}{r}{ 52} &\multicolumn{ 1}{r}{  23:14:08} & \multicolumn{ 1}{r}{+12:31:55} & \multicolumn{ 1}{r}{    89.05} & \multicolumn{ 1}{r}{-43.85} & \multicolumn{ 1}{r}{26.4} & \multicolumn{ 1}{r}{ 27.0 } & \multicolumn{ 1}{r}{0.17} &     -325.8 &     -167.2 &       16.9 &        59.7 &     1310.5 \\

\multicolumn{ 1}{r}{} &\multicolumn{ 1}{r}{} & \multicolumn{ 1}{r}{} & \multicolumn{ 1}{r}{} & \multicolumn{ 1}{r}{} & \multicolumn{ 1}{r}{} & \multicolumn{ 1}{r}{} & \multicolumn{ 1}{r}{} &   -338.4 &     -179.8 &       46.3 &      123.4 &     2710.3 \\
\hline
53 &  22:34:00 &  +12:17:06 &      78.23 &     -38.38 &       17.6 &       12.0 &       0.17 &     -321.3 &     -152.5 &       21.5 &          39.7 &      872.8 \\
 
 \enddata
 \tablenotetext{a} {\scriptsize At the distance of 60 kpc, 1\arcmin = 17pc.}
 \tablenotetext{b} {\scriptsize At the distance of 60 kpc.}
 \label{Stab}
\end{deluxetable}

\clearpage
\begin{deluxetable}{cccccccc}
\rotate
  \tabletypesize{\small}
  \tablewidth{0pc}
  \tablecaption{Mean and Median Values of the Cloud Statistics}
  \tablehead{\colhead{}&\colhead{Unit} &\multicolumn{2}{c}{Complex C}  &\multicolumn{2}{c}{Smoothed C} &\multicolumn{2}{c}{Magellanic Stream}}
  \startdata
     Number of clouds\tablenotemark{a} &&	 \multicolumn{2}{c}{79} &  \multicolumn{2}{c}{16}& \multicolumn{2}{c}{72} \\
     Clouds with single component\tablenotemark{b}  & &  \multicolumn{2}{c}{71} &  \multicolumn{2}{c}{14}& \multicolumn{2}{c}{68} \\
     Single comp. clouds w/ aspect ratio $< $ 1.6\tablenotemark{c} &&\multicolumn{2}{c}{54} &  \multicolumn{2}{c}{13} & \multicolumn{2}{c}{59} \\
     Adopted distance & kpc		& \multicolumn{2}{c}{10} & \multicolumn{2}{c}{10} &  \multicolumn{2}{c}{60} \\
     Velocity range of catalog	& \kms & \multicolumn{2}{c}{-178 - -66} & \multicolumn{2}{c}{-130 - -80} &  \multicolumn{2}{c}{-411 - -307} \\
     \hline
     \colhead{} & &\colhead{Mean} &\colhead{Median} & \colhead{Mean} & \colhead{Median} &\colhead{Mean} &\colhead{Median} \\
     \hline
     $\rm{T_{b,peak}}$\tablenotemark{a} & K			& 0.56&0.41& 0.42& 0.18& 0.33 &0.30 \\ % a
     $\Delta~\rm{V}$\tablenotemark{a}  & \kms			& 26.3& 24.9& 28.2 & 26.2 & 29.3 &27.7 \\ % d
     \vlsr			\tablenotemark{a}&\kms		& -107& -105&-110 &-113 & -367& -364\\ % d
      Mass 			\tablenotemark{a}& \Msun		& 1300&148 &6560 &632 &16400 & 4520\\ %a
     Size			\tablenotemark{a}& \arcmin 		& 35.1&22.2 &90.0 &51.7 &30.4 &22.1 \\ % a
     Size			\tablenotemark{a}& pc			& 102&64.4 &261 &150 & 517& 376\\ %a
     Volume density	\tablenotemark{c}& $\rm{cm^{-3}}$ &0.079 &0.064 &  & \tablenotemark{d}& 0.010&0.0093 \\  %e
     Pressure (P/k)	\tablenotemark{c}& K cm$^{-3}$	&710 & 580&  & \tablenotemark{d}&91& 84\\ %e
     Halo density		\tablenotemark{c}& cm$^{-3}$	& $7.1\times10^{-4}$&$5.8\times10^{-4}$ && \tablenotemark{d}&$9.1\times 10^{-5}$ &$8.4\times 10^{-5}$ \\ %e
     \hline
    \enddata
     \tablenotetext{a} {\scriptsize All the clouds are included. Only the primary component is included for clouds with multiple components.}
     \tablenotetext{b} {\scriptsize Only clouds that are fit with one Gaussian are included.}
      \tablenotetext{c} {\scriptsize Only clouds that are fit with one Gaussian and have aspect ratio (the larger of $\Delta~RA/\Delta Dec$ and $\Delta~Dec/\Delta~RA$) less than 1.6 are included.}
      \tablenotetext{d} {\scriptsize The mean and median values are not derived due to small number of clouds in the smoothed Complex C data.}
      \label{average}
\end{deluxetable}

\clearpage

\clearpage

\clearpage

\clearpage

\clearpage

\clearpage
%Statistics for CXC

\clearpage
%Statistics for CXC

\clearpage
%Statistics for smoothed CXC

\clearpage
%Statistics for smoothed CXC

\clearpage

\clearpage

\clearpage

\clearpage

\end{document}